\documentclass[aps,preprint]{revtex4}
\usepackage[T1]{fontenc}
\usepackage{amsmath,amssymb}   
\usepackage{mathptmx}
\DeclareMathAlphabet{\mathcal}{OMS}{cmsy}{m}{n}
\DeclareSymbolFont{Letters}{OML}{cmm}{m}{it}
\DeclareMathSymbol{\psi}{\mathalpha}{Letters}{32}
\DeclareMathSymbol{\Psi}{\mathalpha}{Letters}{9}
\DeclareMathSymbol{\tau}{\mathalpha}{Letters}{28}
\usepackage{graphicx}  
\usepackage{subfigure}
\usepackage{float}
\usepackage{dcolumn}   
\usepackage{bm}        
\usepackage{booktabs}  
\DeclareMathOperator{\sgn}{sgn}

\DeclareMathOperator{\re}{Re}
\DeclareMathOperator{\im}{Im}
\usepackage[colorlinks,linkcolor=blue,citecolor=blue,urlcolor=blue]{hyperref}
\usepackage{cleveref}
\crefname{equation}{Eq.}{Eqs.}
\crefname{figure}{Fig.}{Figs.}
\crefname{table}{Tab.}{Tabs.}
\crefname{section}{Sec.}{Secs.}
\usepackage{listings}  
\usepackage[dvipsnames, svgnames, x11names]{xcolor} 
\definecolor{IRed}{RGB}{238,99,99}
\definecolor{RBlue}{RGB}{65,105,225}
\usepackage{anyfontsize}
\usepackage{makecell}
\usepackage{multirow}
\usepackage{ulem}

\newcommand{\beq}{\begin{equation}}
\newcommand{\eeq}{\end{equation}}
\newcommand{\ben}{\begin{align}}
\newcommand{\een}{\end{align}}
\newcommand{\bea}{\begin{aligned}}
\newcommand{\eea}{\end{aligned}}
\newcommand{\bes}{\begin{subequations}}
\newcommand{\ees}{\end{subequations}}
\newcommand{\bew}{\begin{widetext}}
\newcommand{\eew}{\end{widetext}}

\numberwithin{equation}{section}
\renewcommand\theequation{\arabic{equation}}

\begin{document}

\title{Quantum Interference Transport in two-dimensional Semi-Dirac Semimetals}
\author{Shihao Bi}
\email{bishihao@stu.scu.edu.cn}
\author{Yiting Deng}
\email{dengyiting@stu.scu.edu.cn}
\author{Yan He}
\email{heyan_ctp@scu.edu.cn}
\author{Peng Li}
\email{lipeng@scu.edu.cn}
\affiliation{College of Physics, Sichuan University, 610064, Chengdu, People's Republic of China}
\affiliation{Key Laboratory of High Energy Density Physics and Technology of Ministry of Education, Sichuan University, 610064,
Chengdu, People's Republic of China}
\date{\today}

\begin{abstract}
Semi-Dirac semimetals have received enthusiastic research both theoretically and experimentally in the recent years.
Due to the anisotropic dispersion, its physical properties are highly direction-dependent. In this work we employ
the Feynman diagrammatic perturbation theory to study the transport properties in quantum diffusive regime. The magneto-conductivity with quantum interference corrections is derived, which demonstrate the weak localization effect in the semi-Dirac semimetal. Furthermore, the origin of anomalous Hall conductivity is also clarified, where both the intrinsic and side-jump contributions vanish and only the skew-scattering gives rise to non-zero transverse conductivity. The conductance fluctuations in both mesoscopic and quantum diffusive regimes are investigated in detail. Our work provides theoretical predictions for transport experiments, which can be examined by conductivity measurements at sufficiently low temperature.
\end{abstract}

\maketitle

\renewcommand\theequation{\Roman{section}.\arabic{equation}}

\section{Introduction}

Semi-Dirac semimetals are exotic phases of matter hosting quasi-particles with linear
and quadratic dispersion relations in different directions, which gives rise to the coalesce of Dirac
fermions and ordinary non-relativistic fermions. Such peculiar dispersion have caused wide research interests
both theoretically and experimentally. A lot of candidate materials such as transition metal oxides
multilayer nano-structures \cite{Pardo2009:PRL,Banerjee2009:PRL,Pardo2010:PRB},
deformed Graphene \cite{Dietl2008:PRL,Montambaux2009:EPJB}, black phosphorus under high
pressure \cite{Xiang2015:PRL,Fei2015:PRB,Gong2016:PRB} and Silicene oxide \cite{Zhong2017:PCCP}, etc., 
have been proposed to realize this dispersion in the past few years. In the meantime, its physical properties 
have been intensively explored, such as Ruderman-Kittel-Kasuya-Yosida (RKKY)
interaction \cite{Duan2017:NJP,Duan2020:arX} between doped magnetic moments,
novel magnetic field dependence of Landau levels \cite{Dietl2008:PRL,Montambaux2009:EPJB,Delplace2010:PRB}
obtained from Wentzel-Kramers-Brillouin (WKB) approximation,
quantum tunneling behavior \cite{Lim2012:PRL,Banerjee2012:PRB,Saha2017:PRB},
transport properties \cite{Adroguer2016:PRB,Park2019:2DM,Ning2020:PRB,Niu2019:PRB,Kim2020:PRB,Narayan2015:PRB},
non-Fermi liquid emergent from long-range Coulomb interaction \cite{Isobe2016:PRL,Gil2016:SR,Kotov2021:PRB}
and its interplay with various types of disorder \cite{Zhao2016:PRB}
in the framework of 2+1 dimensional quantum electrodynamics (QED$_3$),
Floquet dynamics \cite{Narayan2015:PRB,Saha2016:PRB} under the illumination
of polarized off-resonant light, and so on. Among all the physical properties mentioned above,
the transport measurements are most straightforward to perform. However, relatively less attention has been paid
to the transport behaviors of doped semi-Dirac semimetals in the magnetic field.

In realistic materials, impurities may have significant influence on the physical properties
and sometimes are of vital importance. For examples, the impurity effects play an important role in Kondo effect, high temperature superconductors,
quantum Hall insulators, and Anderson localization, etc. When considering the transport properties, it is well-known that
the impurity scattering provides the key mechanism of electron momentum relaxation and leads to finite electrical conductivities.
At low temperature, when the mean free path $\ell_{e}$ is much less than the system size
and phase coherence length $\ell_{\phi}$ (quantum diffusive regime), electrons can maintain their phase coherence even after
being scattered for many times. In this quantum diffusive regime,
the quantum interference between time-reversed scattering loops can give rise to weak localization or anti-localization correction
to the conductivity \cite{Bergmann1984:PR,PALee1985:RMP,Chakravarty1986:PR,HLN1980:PTP,Lu2014:SPIE}. Moreover, impurity scattering
can also generate the anomalous Hall conductivity \cite{Sinitsyn2007:PRB,Nagaosa2010:RMP,Yang2011:PRB,Lu2013:PRB} other than the intrinsic
contribution from the occupied bands.

Motivated by the theoretical and experimental progress on the transport properties of topological materials
in recent years \cite{Lu2011:PRL,Lu2014:PRL,Kim2013:PRL,Xiong2015:Sci,Zhao2016:SR,Li2016:NC,Burkov2014:PRL,Burkov2015:PRB},
we present the study of transport properties of doped semi-Dirac semimetals with quantum interference correction. We will consider the most general form of the impurity scattering mechanism potential. Following the conventional procedure in quantum transport theory, the dominate contributions come from the quantum states near the Fermi surface. We calculate the anisotropic longitudinal conductivities by the linear response theory, in which the effects of quantum interference is considered by including the Cooperon contributions. With finite magnetic field, the magneto-conductivities are obtained by summing over the Landau levels bounded by the mean free path $\ell_e$ and phase coherence length $\ell_{\phi}$. It is also found that the skew-scattering induced by impurity potential can result in non-zero extrinsic anomalous Hall conductivity. Based on the same formalism, we also investigate the conductance fluctuations of semi-Dirac semimetal in both mesoscopic and quantum diffusive regimes.

This paper is organized as follows. The model Hamiltonian and impurity potential are given in \cref{sec:Ham}.
Some notations and physical quantities, such as relaxation times and density of states (DOS) are defined for later convenience. 
In \cref{sec:Cond} we present the diagrammatic perturbation calculation of longitudinal conductivities, which demonstrates the weak localization effect. Next, to understand the suppression of the quantum interference correction by magnetic field, we present the magneto-conductivity formulas with semi-classical quantized Landau levels in \cref{sec:MC}. After that we investigate the anomalous Hall conductivity from side-jump and skew-scattering mechanisms in \cref{sec:AHC}. In \cref{sec:UCF}, we study the universal conductance fluctuation (UCF) of the semi-Dirac semimetals in mesoscopic regime. Finally we draw some conclusions of our main results and make some remarks on possible extensions to our work in \cref{sec:Conc}. For convenience, we set $e=\hbar=1$ throughout the whole paper.

\section{Model Hamiltonian}
\label{sec:Ham}

The Hamiltonian of two dimensional semi-Dirac semimetal is given by
\begin{equation}\begin{aligned}
H(\mathbf{k})=\lambda k_{x}\sigma_{x}+k_{y}^{2}\sigma_{y} ,
\label{eq:ham_v1}
\end{aligned}\end{equation}
Here $\sigma_{x,y,z}$ is the Pauli matrices acting on the pseudospin space, such as the orbital or sublattice degree of freedom, $\lambda$ is the effective velocity in the $x$ direction with the dimension of inverse length, and $\mathbf{k}=(k_{x},k_{y})$ is the two dimensional wave vector. The anisotropic dispersion relation is relativistic in the $x$ direction, and parabolic in the $y$ direction. Its dispersion relation is $\epsilon(\mathbf{k})=\sqrt{\lambda^{2}k_{x}^{2}+k_{y}^{4}}$. The Hamiltonian \cref{eq:ham_v1} violates the time-reversal symmetry 
but possesses the chiral symmetry, and thus belongs to the AIII (unitary) class \cite{Chiu2016:RMP}. 

The impurity potential we considered includes both elastic and pseudospin scattering:
\begin{equation}\begin{aligned}
U(\mathbf{r})=\sum_{\alpha}U_{\alpha}(\mathbf{r})=\sum_{i,\alpha}^{N_{\alpha}} u_{i,\alpha}\sigma_{\alpha}\delta\left(\mathbf{r}-\mathbf{R}_{i}\right),
\end{aligned}\end{equation}
where $\alpha$ runs over $0,x,y$, and $z$. $\sigma_{0}$ is the $2\times2$ identity matrix. The potential $U_{0}(\mathbf{r})$ will cause elastic the scattering on the same pseudospin states, and $U_{x,y}(\mathbf{r})$ will cause the scattering between the different pseudospin states. $U_{z}(\mathbf{r})$
is for the energy splitting between the two pseudospin states. We will assume the scattering of different mechanisms are uncorrelated.
$\mathbf{R}_{i}$s are the locations of the $N_{\alpha}$ randomly distributed impurities, and $u_{i,\alpha}$ is the energy fluctuation
at $\mathbf{R}_{i}$ of $\alpha$-type scattering mechanism with zero mean value and variance $\overline{u_{\alpha}^{2}}$. Therefore, the average
over the impurity configuration of the potential is $\left\langle U(\mathbf{r})\right\rangle _{\mathrm{imp}}=0$, and the potential correlation
is $\left\langle U(\mathbf{r})U(\mathbf{r}')\right\rangle_{\mathrm{imp}}=\sum_{\alpha}n_{\alpha}\overline{u_{\alpha}^{2}}\delta\left(\mathbf{r}-\mathbf{r}'\right)$.
Here the impurity average is
\begin{equation}\begin{aligned}
\left\langle f(\mathbf{r})\right\rangle_{\mathrm{imp}}=\int\prod_{i=1}^{N}\frac{\mathrm{d}\mathbf{R}_{i}}{S}f(\mathbf{r}),
\end{aligned}\end{equation}
$S$ is the sample area and $n_{\alpha}$ is the impurity concentration of $\alpha$-type.

The Bloch wavefunction for the conduction band is
\begin{equation}\begin{aligned}
\left\vert u(\mathbf{k})\right\rangle =\frac{1}{\sqrt{2}}\left[\begin{array}{c}
1\\
\zeta\mathrm{e}^{\mathrm{i}\phi}
\end{array}\right],\; \zeta=\sgn k_{x} ,
\end{aligned}\end{equation}
with $\tan\phi=k_{y}^{2}/\lambda k_{x}$.
The momentum can be parametrized as $k_{x}= \epsilon\cos\phi/\lambda$ and
$k_{y}=\sqrt{\left|\epsilon\sin\phi\right|}\sgn\sin\phi$ near the Fermi surface. The $\sgn(x)$ is the sign function.
Then we can define the velocity operator $v_{\mathbf{k}}^{\alpha}=\dfrac{\mathrm{d}\epsilon}{\mathrm{d}k_{\alpha}}$
in the $x$ and $y$ direction, respectively.
\begin{equation}\begin{aligned}
v_{\mathbf{k}}^{x}=& \lambda \cos\phi , \\
v_{\mathbf{k}}^{y}=& 2\sqrt{\left|\epsilon\sin\phi\right|}\sin\phi .
\end{aligned}\end{equation}

The density of states (DOS) near the Fermi surface is $\rho(\epsilon)=\dfrac{K(1/2)}{\sqrt{2}\lambda\pi^{2}}\epsilon^{1/2}$, obtained
after the integration of the angular variable
\begin{equation}\begin{aligned}
\int\mathrm{d}\Omega=\int_{0}^{2\pi}\frac{\mathrm{d}\phi}{4K(1/2)\sqrt{2\left|\sin\phi\right|}} .
\end{aligned}\end{equation}
$K(x)$ is the complete elliptic integral of the first kind, and $K(1/2) \approx 1.854$. We see that the DOS has a square-root
dependence of the Fermi energy, which is distinguished from the linear dependence for Graphene or constant for two-dimensional
electron gas. Here and throughout the text we assume that the Fermi energy is positive and lies across the conduction band.
Finally, the relaxation time near the Fermi surface can be obtained by the Fermi golden rule
\begin{equation}\begin{aligned}
\tau^{-1}=&2\pi\sum_{\mathbf{k}'}\left\langle \big|U_{\mathbf{k}\mathbf{k}'}\big|^{2}\right\rangle _{\mathrm{imp}}\delta\left(\epsilon_{\mathbf{k}'}-\epsilon_{F}\right) \\
=&2\pi\rho(\epsilon_{F}) \int\mathrm{d}\Omega \sum_{\alpha}n_{\alpha}\overline{u_{\alpha}^{2}}\left|A_{\alpha,\mathbf{k}\mathbf{k}'}\right|^{2} \\
=&\pi\rho(\epsilon_{F})\sum_{\alpha}n_{\alpha}\overline{u_{\alpha}^{2}}
\end{aligned}\end{equation}
Here $U_{\mathbf{k}\mathbf{k}'}$ is the Born scattering amplitude between two momenta,
\begin{equation}\begin{aligned}
U_{\mathbf{k}\mathbf{k}'}=\int\frac{\mathrm{d}^{2}\mathbf{r}}{S}\sum_{i,\alpha}u_{i,\alpha}A_{\alpha,\mathbf{k}\mathbf{k}'}\delta\left(\mathbf{r}-\mathbf{R}_{i}\right)\mathrm{e}^{-\mathrm{i}\left(\mathbf{k}-\mathbf{k}'\right)\cdot\mathbf{r}}
\end{aligned}\end{equation}
and $A_{\alpha,\mathbf{k}\mathbf{k}'}=\left\langle u(\mathbf{k})\right|\sigma_{\alpha}\left|u(\mathbf{k}')\right\rangle$.
We find that for $\alpha$-type scattering the relaxation time
$\tau^{-1}_{\alpha}=\pi\rho(\epsilon_{F}) n_{\alpha}\overline{u_{\alpha}^{2}}$, thus we have $\tau^{-1}=\sum_{\alpha}\tau^{-1}_{\alpha}$. In the whole paper, we will set $\tau_{x}=\tau_{y}$ as they both denote the relaxation time of scattering between different pseudospin states.
\begin{figure}
\centering
\includegraphics[width=14cm]{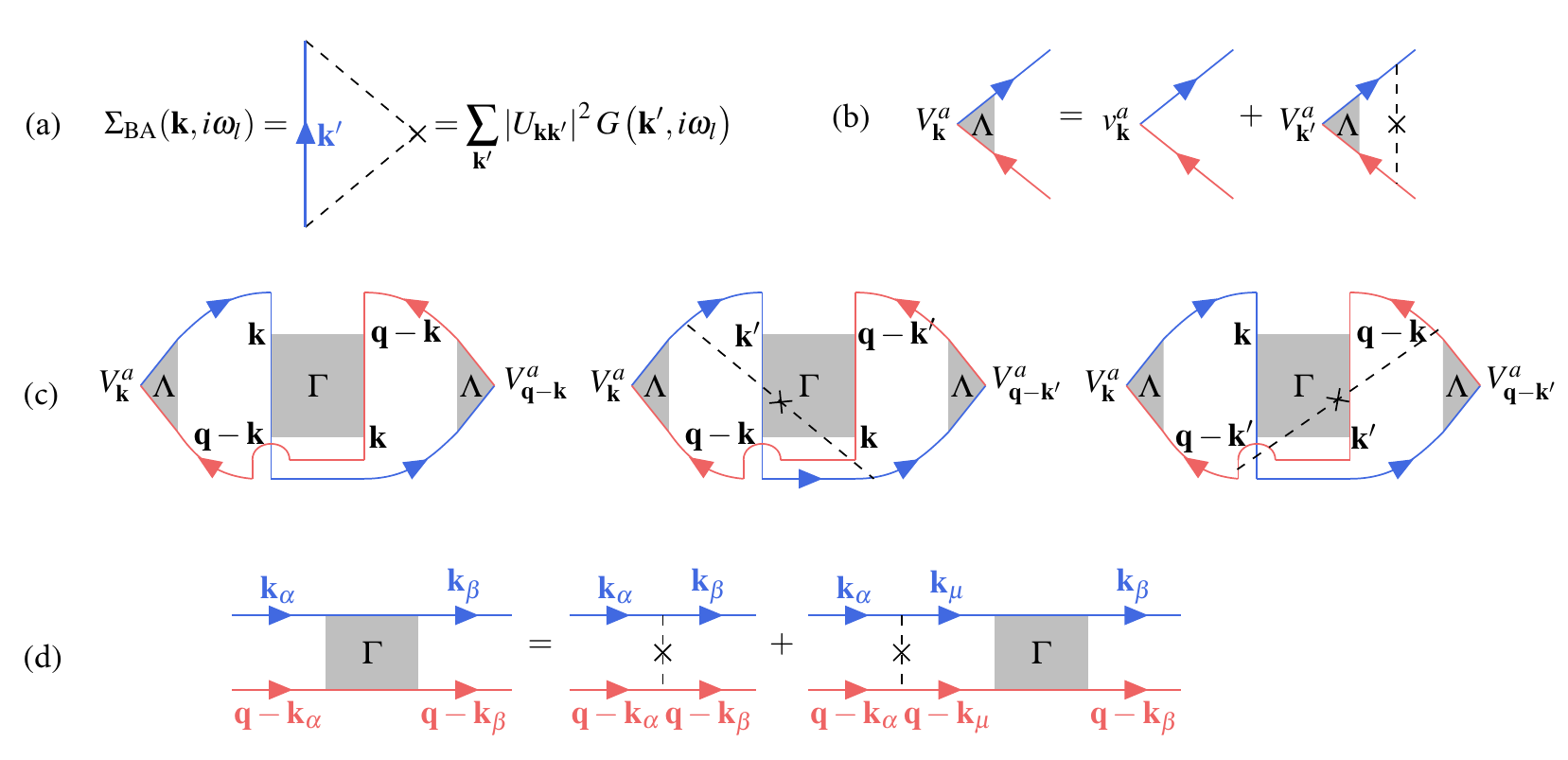}
\caption{Feynman diagrams for the conductivity calculation.
(a) Born approximation for the disorder-averaged self energy.
The cross is the impurity and the dashed lines are the impurity potential.
(b) Vertex correction to the velocity at the Boltzmann level.
(c) Quantum conductivity correction from the bare and two dressed Hikami boxes.
(d) The Bethe-Salpeter equation for the Cooperon diagram.}
\label{fig:LC}
\end{figure}

\section{Diagrammatic Perturbation Approach to the Conductivity}
\label{sec:Cond}

To compute the electrical conductivity we employ the Feynman diagrammatic perturbation method based on the
Matsubara Green's function. The necessary diagrams are listed in \cref{fig:LC}. The Matsubara Green's
function under the Born approximation of the disorder scattering has the form
\begin{equation}\begin{aligned}
G(\mathbf{k},\mathrm{i}\omega_{m})=\frac{1}{\mathrm{i}\omega_{m}-\epsilon_{\mathbf{k}}+\mathrm{i}/(2\tau)\sgn\omega_{m}}
\label{eq:MGF}
\end{aligned}\end{equation}
where $\omega_{m}=\left(2m+1\right)\pi/\beta$ is the fermionic Matsubara frequency, and $\beta$ is the inverse temperature.
After the calculation we will make analytic continuation $\mathrm{i}\nu_{n}\to\nu+0^{+}$ to obtain the zero temperature conductivity.

\subsection{Drude-Boltzmann Conductivity}

The Drude conductivity is given by the simple bubble diagram with bare velocity vertex as follows
\begin{equation}\begin{aligned}
\sigma_{\mu\mu}(\mathbf{q},\mathrm{i}\nu_{n})
=&\frac{1}{\beta\nu_{n}}\sum_{m}\int\frac{\mathrm{d}^{2}\mathbf{k}}{\left(2\pi\right)^{2}}
v_{\mathbf{k}}^{\mu}G(\mathbf{k},\mathrm{i}\omega_{m})v_{\mathbf{k+q}}^{\mu}G(\mathbf{k+q},\mathrm{i}\omega_{m}+\mathrm{i}\nu_{n})
\end{aligned}\end{equation}
here $\mathbf{q}$ is the momentum transferred, $\nu_{n}$ is the bosonic Matsubara frequency and $v_{\mathbf{k}}^\mu$ is the velocity operator defined in last section.
In the long wavelength limit $\mathbf{q} \to 0$, the zero temperature Drude conductivities
contributed from the electrons near the Fermi surface are
\begin{equation}\begin{aligned}
&\sigma_{xx}(0)=\dfrac{2}{3}\rho(\epsilon_{F}) \lambda^{2} \tau,\\
&\sigma_{yy}(0)=\dfrac{6\pi}{5K(1/2)^{2}}\rho(\epsilon_{F}) \left|\epsilon_{F}\right| \tau
\end{aligned}\end{equation}
According to the Einstein relation $\sigma_{\mu\mu}=D_{\mu}\rho(\epsilon_{F})$, we find that the bare diffusion constants are
\begin{equation}\begin{aligned}
&D_{x}=2\lambda^{2}\tau/3,\\
&D_{y}=\dfrac{6\pi}{5K(1/2)^{2}} \left|\epsilon_{F}\right| \tau
\label{eq:Dxy}
\end{aligned}\end{equation}

In order to find results consistent with Boltzmann equation, one has to take into account the ladder diagram shown in \cref{fig:LC} (b),
then the velocity operator is re-normalized as
\begin{equation}\begin{aligned}
V_{\mathbf{k}}^{\mu}=v_{\mathbf{k}}^{\mu}+\sum_{\mathbf{k}'}\left\langle \big|U_{\mathbf{k}\mathbf{k}'}\big|^{2}\right\rangle_{\mathrm{imp}}G(\mathbf{k},\mathrm{i}\omega_{m})V_{\mathbf{k}'}^{\mu}G(\mathbf{k+q},\mathrm{i}\omega_{m}+\mathrm{i}\nu_{n}) .
\end{aligned}\end{equation}
If we assume that the trial solution is $V_{\mathbf{k}}^{\mu}=\eta_{\mu}v_{\mathbf{k}}^{\mu}$, then we can determine that the vertex
correction coefficient $\eta_{\mu}$ at zero temperature is
\begin{equation}\begin{aligned}
1-\eta_{\mu}^{-1}=2\pi\rho(\epsilon_{F})\tau \dfrac{\int\mathrm{d}\Omega\mathrm{d}\Omega'v_{\mathbf{k}}^{\mu}v_{\mathbf{k}'}^{\mu}\sum_{\alpha}n_{\alpha}\overline{u_{\alpha}^{2}}\left|A_{\alpha,\mathbf{k}\mathbf{k}'}\right|^{2}}{\int\mathrm{d}\Omega v_{\mathbf{k}}^{\mu}v_{\mathbf{k}}^{\mu}} .
\end{aligned}\end{equation}
After performing the integral, we find that the vertex correction coefficients are
\begin{equation}\begin{aligned}
\eta_{x}^{-1}=&1-\dfrac{2}{3}\left(\frac{\tau}{\tau_{0}}-\frac{\tau}{\tau_{z}}\right) , \\
\eta_{y}^{-1}=&1-\frac{5\pi}{48}\left(\frac{\tau}{\tau_{0}}-\frac{\tau}{\tau_{z}}\right) .
\end{aligned}\end{equation}
And the Boltzmann conductivity can be obtained with one dressed velocity operator inserted in the bubble diagram as
\begin{equation}\begin{aligned}
\sigma_{\mu\mu}(\mathbf{q},\mathrm{i}\nu_{n})
=&\frac{1}{\beta\nu_{n}}\sum_{m}\int\frac{\mathrm{d}^{2}\mathbf{k}}{\left(2\pi\right)^{2}}
V_{\mathbf{k}}^{\mu}G(\mathbf{k},\mathrm{i}\omega_{m})v_{\mathbf{k+q}}^{\mu}G(\mathbf{k+q},\mathrm{i}\omega_{m}+\mathrm{i}\nu_{n})
\end{aligned}\end{equation}
Then it is easy to see that the conductivities are given by
\begin{equation}\begin{aligned}
&\sigma_{xx}(0)=\dfrac{2}{3}\rho(\epsilon_{F}) \lambda^{2} \tau \eta_{x},\\
&\sigma_{yy}(0)=\dfrac{6\pi}{5K(1/2)^{2}}\rho(\epsilon_{F}) \left|\epsilon_{F}\right| \tau \eta_{y}
\end{aligned}\end{equation}
The corresponding corrected diffusion constants are $\mathcal{D}_{\mu}=\eta_{\mu}D_{\mu}$.
It is known that the semi-classical conductivity will not response to the weak magnetic field.
To reveal the influence of magnetic field on the conductivity, we should consider the quantum interference
between the time-reversal paths, which correspond to the maximally crossed diagrams in \cref{fig:LC} (c,d).

\subsection{Quantum Interference Correction}

The quantum interference correction is given by the maximally crossed diagram with one bare and two dressed Hikami boxes.
The conductivity with one bare Hikami box contribution is
\begin{equation}\begin{aligned}
\sigma_{\mu\mu}^{0}(\mathbf{q},\mathrm{i}\nu_{n})=&\frac{1}{\beta\nu_{n}}\sum_{m}\int\frac{\mathrm{d}^{2}\mathbf{k}}{\left(2\pi\right)^{2}}
\sum_{\mathbf{q}}V_{\mathbf{k}}^{\mu}V_{\mathbf{q}-\mathbf{k}}^{\mu} G\left(\mathbf{k},\mathrm{i}\omega_{m}\right)G\left(\mathbf{q}-\mathbf{k},\mathrm{i}\omega_{m}\right) \\
& \times \Gamma(\mathbf{q},\mathrm{i}\nu_{n})G\left(\mathbf{q}-\mathbf{k},\mathrm{i}\omega_{m}+\mathrm{i}\nu_{n}\right)G\left(\mathbf{k},\mathrm{i}\omega_{m}+\mathrm{i}\nu_{n}\right)
\label{sig-H0}
\end{aligned}\end{equation}
At $T=0$, one can arrive at the following result
\begin{equation}\begin{aligned}
\sigma_{\mu\mu}^{0}(0)=&-2D_{\mu}\rho(\epsilon_{F})\tau^{2}\eta_{x}^{2}\sum_{\mathbf{q}}\Gamma(\mathbf{q})
\end{aligned}\end{equation}
with Cooperon vertex function $\Gamma(\mathbf{q})$ to be determined later.

The two dressed Hikami boxes can be written as follows
\begin{equation}\begin{aligned}
\sigma_{\mu\mu}^{R}(\mathbf{q},\mathrm{i}\nu_{n})=&\frac{1}{\beta\nu_{n}}\sum_{m}\int\frac{\mathrm{d}^{2}\mathbf{k}}{\left(2\pi\right)^{2}}\int\frac{\mathrm{d}^{2}\mathbf{k}'}{\left(2\pi\right)^{2}}\sum_{\mathbf{q}}\left\langle U_{\mathbf{k}'\mathbf{k}}U_{\mathbf{q-k}'\mathbf{q-k}}\right\rangle_{\mathrm{imp}}
V_{\mathbf{k}}^{\mu}V_{\mathbf{q}-\mathbf{k}'}^{\mu}G\left(\mathbf{k},\mathrm{i}\omega_{m}\right)G\left(\mathbf{k}',\mathrm{i}\omega_{m}\right) \\
&\times G\left(\mathbf{q}-\mathbf{k},\mathrm{i}\omega_{m}\right)G\left(\mathbf{q}-\mathbf{k}',\mathrm{i}\omega_{m}\right)\Gamma(\mathbf{q},\mathrm{i}\nu_{n})G\left(\mathbf{q}-\mathbf{k}',\mathrm{i}\omega_{m}+\mathrm{i}\nu_{n}\right)G\left(\mathbf{k},\mathrm{i}\omega_{m}+\mathrm{i}\nu_{n}\right) \\
\sigma_{\mu\mu}^{A}(\mathbf{q},\mathrm{i}\nu_{n})=&\frac{1}{\beta\nu_{n}}\sum_{m}\int\frac{\mathrm{d}^{2}\mathbf{k}}{\left(2\pi\right)^{2}}\int\frac{\mathrm{d}^{2}\mathbf{k}'}{\left(2\pi\right)^{2}}\sum_{\mathbf{q}}\left\langle U_{\mathbf{k}'\mathbf{k}}U_{\mathbf{q-k}'\mathbf{q-k}}\right\rangle_{\mathrm{imp}}
V_{\mathbf{k}}^{\mu}V_{\mathbf{q}-\mathbf{k}'}^{\mu}G\left(\mathbf{k},\mathrm{i}\omega_{m}\right)G\left(\mathbf{q}-\mathbf{k}',\mathrm{i}\omega_{m}\right) \\
&\times\Gamma(\mathbf{q},\mathrm{i}\nu_{n})G\left(\mathbf{k},\mathrm{i}\omega_{m}+\mathrm{i}\nu_{n}\right)G\left(\mathbf{k}',\mathrm{i}\omega_{m}+\mathrm{i}\nu_{n}\right)G\left(\mathbf{q}-\mathbf{k},\mathrm{i}\omega_{m}+\mathrm{i}\nu_{n}\right)G\left(\mathbf{q}-\mathbf{k}',\mathrm{i}\omega_{m}+\mathrm{i}\nu_{n}\right)
\label{sig-H1}
\end{aligned}\end{equation}
Despite the the above complicated expressions, we find that the two dressed Hikami boxes contribute equally and are proportional to the bare Hikami box contributions. The proportional factors can be computed to give relatively simple results as follows
\begin{equation}\begin{aligned}
\xi_{x}=&\frac{\sigma_{xx}^{R/A}(0)}{\sigma_{xx}^{0}(0)}
=\frac{\pi\rho(\epsilon_{F})\tau\int\mathrm{d}\Omega\mathrm{d}\Omega'v_{\mathbf{k}}^{x}v_{-\mathbf{k}'}^{x}\left\langle U_{\mathbf{k}'\mathbf{k}}U_{\mathbf{-k}'\mathbf{-k}}\right\rangle _{\mathrm{imp}}}{\int\mathrm{d}\Omega v_{\mathbf{k}}^{x}v_{\mathbf{k}}^{x}}
=-\frac{1}{3}\left(\frac{\tau}{\tau_{0}}-\frac{\tau}{\tau_{z}}-2\frac{\tau}{\tau_{x}}\right) \\
\xi_{y}=&\frac{\sigma_{yy}^{R/A}(0)}{\sigma_{yy}^{0}(0)}
=\frac{\pi\rho(\epsilon_{F})\tau\int\mathrm{d}\Omega\mathrm{d}\Omega'v_{\mathbf{k}}^{y}v_{-\mathbf{k}'}^{y}\left\langle U_{\mathbf{k}'\mathbf{k}}U_{\mathbf{-k}'\mathbf{-k}}\right\rangle _{\mathrm{imp}}}{\int\mathrm{d}\Omega v_{\mathbf{k}}^{y}v_{\mathbf{k}}^{y}}
=-\frac{5\pi}{96}\left(\frac{\tau}{\tau_{0}}+\frac{\tau}{\tau_{z}}\right)
\end{aligned}\end{equation}

Put all the above results together, we find that the quantum interference correction to the conductivity is
\begin{equation}\begin{aligned}
\sigma_{\mu\mu}^{\mathrm{qi}}=-2D_{\mu}\rho(\epsilon_{F})\tau\eta_{\mu}^{2}\left(1+2\xi_{\mu}\right)\sum_{\mathbf{q}}\Gamma(\mathbf{q})
\label{eq:sig-qi}
\end{aligned}\end{equation}
In order to find the final results of conductivity, we will compute the Cooperon vertex function $\Gamma(\mathbf{q})$ in the next sub-section.

\subsection{Bethe-Salpeter Equation for the Cooperon}

We start from the bare Cooperon vertex function which is given by
\begin{equation}\begin{aligned}
\Gamma_{\mathbf{k}\mathbf{k}'}^{0}  =\left\langle U_{\mathbf{k}'\mathbf{k}}U_{\mathbf{-k}'\mathbf{-k}}\right\rangle _{\mathrm{imp}}
&=\frac{1}{2\pi\rho\tau S}\left[\left(\frac{\tau}{\tau_{0}}-\frac{\tau}{\tau_{z}}\right)+\left(\frac{\tau}{\tau_{0}}+\frac{\tau}{\tau_{z}}\right)\cos\left(\phi-\phi'\right)-2\frac{\tau}{\tau_{x}}\cos\left(\phi+\phi'\right)\right]
\end{aligned}\end{equation}
It is more convenient to rewrite $\Gamma^0$ as the following form
\begin{equation}\begin{aligned}
\Gamma_{\mathbf{k}\mathbf{k}'}^{0}=\frac{1}{2\pi\rho\tau S}\sum_{ab} z_{ab}\mathrm{e}^{\mathrm{i}(a\phi-b\phi')}
\label{eq:ga0}
\end{aligned}\end{equation}
where $a,b\in\left\{ -1,0,1\right\} $. The non-zero components of coefficients $z_{ab}$ are
\begin{equation}\begin{aligned}
&z_{00}=\frac{\tau}{\tau_{0}}-\frac{\tau}{\tau_{z}} \\
&z_{11}=z_{\overline{1}\overline{1}}=\frac{1}{2}\left(\frac{\tau}{\tau_{0}}+\frac{\tau}{\tau_{z}}\right) \\
&z_{1\overline{1}}=z_{\overline{1}1}=-\frac{\tau}{\tau_{x}}
\end{aligned}\end{equation}
Here we denote $\overline{1}=-1$.
Inspired by this, we make an ansatz of the full Cooperon vertex function with the same function form
\begin{equation}\begin{aligned}
\Gamma_{\mathbf{k}\mathbf{k}'}=\frac{1}{2\pi\rho\tau S}\sum_{ab} Z_{ab}\mathrm{e}^{\mathrm{i}(a\phi-b\phi')}
\label{eq:ga}
\end{aligned}\end{equation}
Here the coefficients $Z_{ab}$ are yet to be determined.

The Bethe-Salpeter equation for the Cooperon vertex function is
\begin{equation}\begin{aligned}
\Gamma_{\mathbf{k}_{\alpha}\mathbf{k}_{\beta}}=&\Gamma_{\mathbf{k}_{\alpha}\mathbf{k}_{\beta}}^{0}+\sum_{\mathbf{k}_{\mu}}\Gamma_{\mathbf{k}_{\alpha}\mathbf{k}_{\mu}}^{0}G\left(\mathbf{k}_{\mu},\mathrm{i}\omega_{m}\right)G\left(\mathbf{q}-\mathbf{k}_{\mu},\mathrm{i}\omega_{m}+\mathrm{i}\nu_{n}\right)\Gamma_{\mathbf{k}_{\mu}\mathbf{k}_{\beta}} \\
\end{aligned}\end{equation}
Plug in Eq.(\ref{eq:ga0}) and (\ref{eq:ga}) and performing the of the momentum summation of the electron propagator, we find that the Bethe-Salpeter equation transfers to an equation $Z_{ab}$
\begin{equation}\begin{aligned}
Z_{ab}=&z_{ab}+\sum_{cd} z_{ac}\int\frac{\mathrm{d}\Omega_{\mu}}{1+\mathrm{i}\mathbf{v_{\mu}\cdot q}\tau}\mathrm{e}^{-\mathrm{i}(c-d)\phi_{\mu}}Z_{db}
\end{aligned}\end{equation}
Here $\mathbf{v}_{\mu}=\dfrac{\partial\epsilon_{\mathbf{k}}}{\partial\mathbf{k}_{\mu}}=\left(\lambda\cos\phi_{\mu},2\sqrt{\left|\epsilon\sin\phi_{\mu}\right|}\sin\phi_{\mu}\right)$
and $\mathbf{v_{\mu}\cdot q}=\lambda q_{x}\cos\phi_{\mu}+2q_{y}\sqrt{\left|\epsilon\sin\phi_{\mu}\right|}\sin\phi_{\mu}$.
We can further simplify the equation by expanding the fraction in the integrand to the $q^{2}$ term and define
\begin{equation}\begin{aligned}
\Phi_{cd}=\int\mathrm{d}\Omega_{\mu}\left[1-\mathrm{i}\mathbf{v_{\mu}\cdot q}\tau-\left(\mathbf{v_{\mu}\cdot q}\tau\right)^{2}\right]\mathrm{e}^{-\mathrm{i}\left(c-d\right)\phi_{\mu}}
\end{aligned}\end{equation}
Then the Bethe-Salpeter equation reduces to a simple matrix equation $\mathbf{Z}=\mathbf{z}+\mathbf{z}\bm{\Phi}\mathbf{Z}$, and its solution is
\begin{equation}\begin{aligned}
\mathbf{Z}=&\left(1-\mathbf{z}\bm{\Phi}\right)^{-1}\mathbf{z}
\end{aligned}\end{equation}

We will only keep the most divergent term which is given as
\begin{equation}\begin{aligned}
&Z_{00}\approx\dfrac{1}{\left(\dfrac{1}{z_{00}}-1\right)+g_{x}D_{x}q_{x}^{2}\tau+g_{y}D_{y}q_{y}^{2}\tau}\\
&g_x=\left(1+2\dfrac{z_{11}}{z_{00}}+\dfrac{10}{7}\dfrac{z_{1\overline{1}}}{z_{00}}\right),\quad
g_y=\left(1+2\dfrac{z_{11}}{z_{00}}-\dfrac{10}{9}\dfrac{z_{1\overline{1}}}{z_{00}}\right)
\label{eq:Z00}
\end{aligned}\end{equation}
Here $D_{x,y}$ are bare diffusion constants in Eq.(\ref{eq:Dxy}).

Make use the above result, we find that the summation of the Cooperon vertex function is
\begin{equation}\begin{aligned}
\sum_{\mathbf{q}}\Gamma(\mathbf{q}) 
=&\frac{1}{8\pi^{2}\rho(\epsilon_{F})\tau^{2}\sqrt{g_{x}g_{y}D_{x}D_{y}}}\ln\frac{\ell_{0}^{-2}+\ell_{e}^{-2}}{\ell_{0}^{-2}+\ell_{\phi}^{-2}}
\end{aligned}\end{equation}
Here we define a combined length parameter $\ell_{0}^{-2}=\left(\dfrac{1}{z_{00}}-1\right)\left(g_{x}D_{x}\tau\right)^{-1}$, and the momentum summation is bounded by the mean free path $\ell_{e}$ and the phase coherence length $\ell_{\phi}$.

Substitute the Cooperon vertex function into Eq.(\ref{eq:sig-qi}), we find the conductivity with the quantum interference correction as
\begin{equation}\begin{aligned}
\sigma_{\mu\mu}^{\mathrm{qi}}=-\frac{e^{2}}{\hbar} \frac{D_{\mu}\eta_{\mu}^{2}\left(1+2\xi_{\mu}\right)}{4\pi^{2}\sqrt{g_{x}g_{y}D_{x}D_{y}}}
\ln\frac{\ell_{0}^{-2}+\ell_{e}^{-2}}{\ell_{0}^{-2}+\ell_{\phi}^{-2}}
\label{eq:sig-B0}
\end{aligned}\end{equation}
which displays the \textit{weak localization} effects. In the above expression, we have restored the unit of universal conductance $e^{2}/\hbar$. When the magnetic field is applied, the quantum interference effect will be suppressed, and the conductivity is expected to increase. We will turn to the effects of magnetic field in the next section.

\section{Magneto-conductivity}
\label{sec:MC}

Applying a magnetic field along the $z$ direction, we expect the motion on the $xy$ plane will become quantized as the ordinary free fermions. The vector potential in the Landau gauge is $\mathbf{A}=\left(0,Bx,0\right)$.
By making the Peierls substitution, the Hamiltonian \cref{eq:ham_v1} becomes
\begin{equation}\begin{aligned}
H(\mathbf{k}+e\mathbf{A})=\lambda k_{x}\sigma_{x}+\left(k_{y}+eBx\right)^{2}\sigma_{y} .
\label{eq:ham_v2}
\end{aligned}\end{equation}
The eigenvalue equation for the two-components wavefunction $\psi$ is $H\psi=E\psi$, which is also equivalent to
a second-order differential equation as \cite{Dietl2008:PRL,Montambaux2009:EPJB}
\begin{equation}\begin{aligned}
\left(-\lambda^{2}\partial_{x}^{2}+\mathrm{i}\lambda e^{2}B^{2}\left[-\mathrm{i}\partial_{x},x^{2}\right]\sigma_{z}+e^{4}B^{4}x^{4}\right)\psi=E^{2}\psi .
\label{eq:EigenEqs}
\end{aligned}\end{equation}
Clearly, the $x^4$ term will dominate for large $x$, thus we can drop the commutator
to simplify the above equation. Then we can find the quantized eigenvalues by the WKB quantization condition
\begin{equation}\begin{aligned}
\int p\mathrm{d}x=\left(n+\frac{1}{2}\right)\pi .
\end{aligned}\end{equation}
After finish the integral, we find that
\begin{equation}\begin{aligned}
E_{n}=\gamma\left(\frac{\lambda}{\ell_{B}^{2}}\right)^{2/3}\left(n+\frac{1}{2}\right)^{2/3} \;,
\gamma=\left[\frac{3\pi\sqrt{2\pi}}{\Gamma\left(1/4\right)^{2}}\right]^{2/3}\approx1.478 ,
\end{aligned}\end{equation}
Here the magnetic length is $\ell_{B}=\sqrt{\hbar/eB}$.
Because of this Landau level quantization, in the Cooperon vertex function, the momentum $\mathbf{q}$ is then quantized into
$q_{n} = 2\gamma\left(\lambda\ell_{B}^{4}\right)^{-1/3}\left(n+\dfrac{1}{2}\right)^{2/3}$.

To obtain the conductivity correction formula in a finite magnetic field $B$, we insert a
Dirac $\delta$ function in the Cooperon vertex function $\Gamma(\mathbf{q})$ to restrict the momentum to quantized values:
\begin{equation}\begin{aligned}
\sum_{\mathbf{q}}\Gamma(\mathbf{q})\Rightarrow&
\sum_{\mathbf{q}}\Gamma(\mathbf{q})
\times \sum_{n} \delta\left[n+\frac{1}{2}-\left(2\gamma\right)^{-3/2}\sqrt{\lambda}\ell_{B}^{2}q^{3/2}\right]
\label{eq:cp_B}
\end{aligned}\end{equation}
Since the momentum summation is bounded as $\ell_e^{-1}<q<\ell_\phi^{-1}$, correspondingly, the lower and upper bound of the Landau Levels, $n_{\mathrm{L}}$ and $n_{\mathrm{U}}$,
cut-off by the mean free path $\ell_{e}$ and the phase coherence length $\ell_{\phi}$.
\begin{equation}
n_{\mathrm{L}}=\chi\ell_{\phi}^{-3/2}-\frac12,\quad
n_{\mathrm{U}}=\chi\ell_{e}^{-3/2}-\frac12
\end{equation}
Here we define the abbreviation $\chi=\left(2\gamma\right)^{-3/2}\sqrt{\lambda}\ell_{B}^{2}$.

To carry out the Landau level summation, we first consider a simpler case with $\ell_{0}^{-2}=0$. In this case, \cref{eq:cp_B} simply reduce to a harmonic series and we find
\begin{equation}\begin{aligned}
\sigma_{\mu\mu}^{\mathrm{qi}}(B,\ell_{0}^{-2}=0)=-\frac{D_{\mu}\eta_{\mu}^{2}\left(1+2\xi_{\mu}\right)}{3\pi^{2}\sqrt{g_{x}g_{y}D_{x}D_{y}}}
\left[\psi\left(\chi\ell_{e}^{-3/2}+\frac{1}{2}\right)-\psi\left(\chi\ell_{\phi}^{-3/2}+\frac{1}{2}\right)\right]
\label{eq:c_qi_B_v1}
\end{aligned}\end{equation}
If the magnet field $B$ is very small, then the magnetic length $\ell_{B}\to+\infty$. By making use of the asymptotic
behavior of digamma function $\psi\left(x+\dfrac{1}{2}\right)\approx\ln x$, one finds
\begin{equation}\begin{aligned}
\psi\left(\chi\ell_{e}^{-3/2}+\frac{1}{2}\right)-\psi\left(\chi\ell_{\phi}^{-3/2}+\frac{1}{2}\right)=\frac{3}{4}\ln\frac{\ell_{e}^{-2}}{\ell_{\phi}^{-2}} .
\end{aligned}\end{equation}
which reproduce Eq.(\ref{eq:sig-B0}) with $\ell_{0}^{-2}=0$.

For non-zero $\ell_{0}^{-2}$, we simply make the replacement $\ell_{e/\phi}^{-2}\to\ell_{e/\phi}^{-2}+\ell_{0}^{-2}$ in \cref{eq:c_qi_B_v1} and obtain
\begin{equation}\begin{aligned}
\sigma_{\mu\mu}^{\mathrm{qi}}(B)=-\frac{D_{\mu}\eta_{\mu}^{2}\left(1+2\xi_{\mu}\right)}{3\pi^{2}\sqrt{g_{x}g_{y}D_{x}D_{y}}}
\left[\psi\left(\chi\left(\ell_{e}^{-2}+\ell_{0}^{-2}\right)^{3/4}+\frac{1}{2}\right)-\psi\left(\chi\left(\ell_{\phi}^{-2}+\ell_{0}^{-2}\right)^{3/4}+\frac{1}{2}\right)\right]
\label{eq:c_qi_B_v2}
\end{aligned}\end{equation}

\begin{figure}
\centering
\subfigure{\begin{minipage}[t]{0.45\textwidth}
\centering
\includegraphics[scale=0.8]{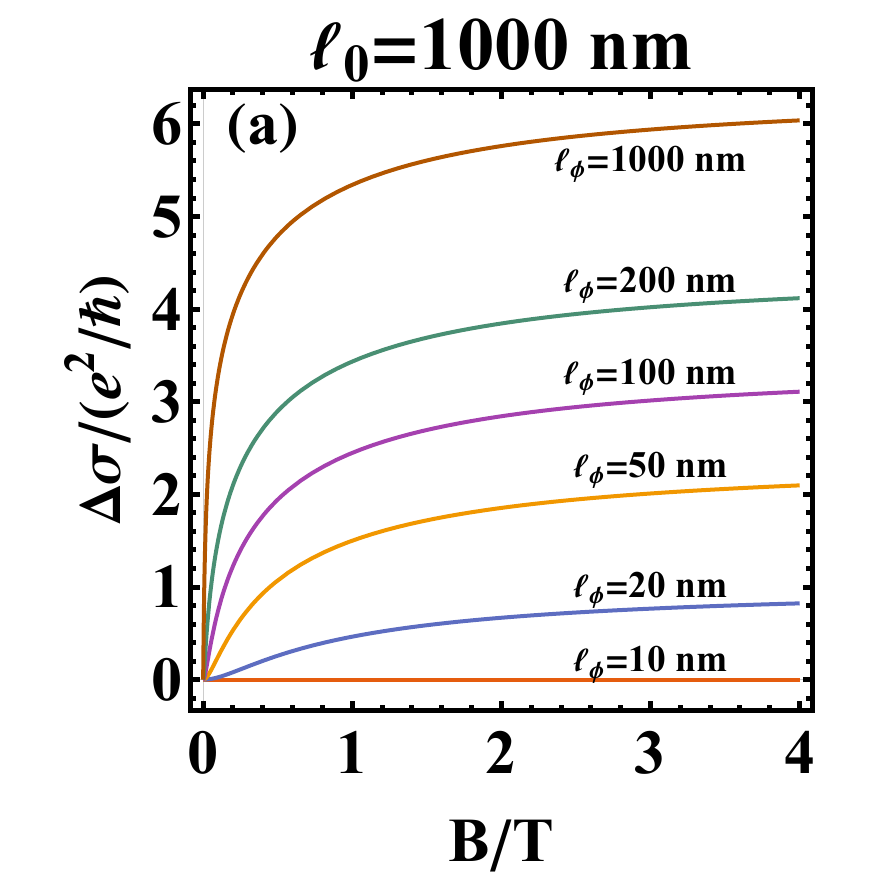}
\end{minipage}
}
\subfigure{\begin{minipage}[t]{0.45\textwidth}
\centering
\includegraphics[scale=0.8]{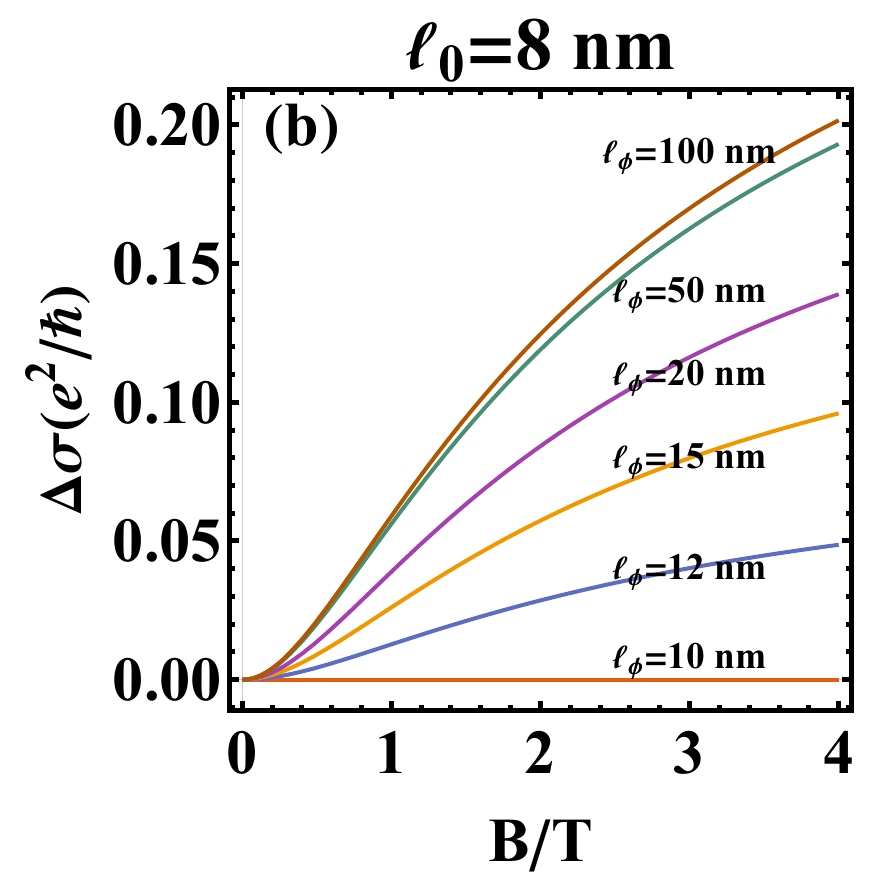}
\end{minipage}
}
\caption{Magneto-conductivity versus magnetic field strength. In panel (a), the adjustable $\ell_{0}=1000$nm and the phase coherence length $\ell_\phi=$10, 20, 50, 100, 200, 1000nm from bottom to top. In panel (b), the adjustable $\ell_{0}=8$nm and the phase coherence length $\ell_\phi=$10, 12, 20, 50, 200nm from bottom to top. For both panles, the mean free path $\ell_{e}=10$nm. The effective velocity $\lambda$ is the inverse of the lattice constant $a^{-1}$ with $a=0.246$ nm as in Graphene. The magnetic length taken is to be $\ell_{B}$=25.66 nm/$\sqrt{B}$, with $B$ in Tesla.}
\label{fig:MC}
\end{figure}

The magneto-conductivity is the change of conductivity induced by the applied magnetic field, which is defined as $\Delta\sigma(B)=\sigma_{\mu\mu}^{\mathrm{qi}}(B)-\sigma_{\mu\mu}^{\mathrm{qi}}(0)$.
To find a qualitative picture of above results, we drop the prefactor before the large brackets of \cref{eq:c_qi_B_v2}, and choose some parameters of interest to show the magnetic field dependence of magneto-conductivity in \cref{fig:MC}. The effective velocity in the $x$-direction is proportional to the inverse of lattice constant $a$, and we take $a=0.246$ nm as in the Graphene. The mean free path is set as 10 nm. The phase coherence length is of order 100 nm typically,
and can be tuned by the temperature in the experiment.

In \cref{fig:MC}, the magneto-conductivity is displayed  in the two limiting cases where the adjustable length parameter $\ell_{0}=1000$nm in panel (a) and  $\ell_{0}=8$nm in panel (b). In both panels, we take a series of values for $\ell_{\phi}$ and plot $\Delta\sigma$ as a function of $B$. The panel (a) corresponds to the case where the $U_{x,y,z}(\mathbf{r})$ impurity potentials are absent, which means $\tau=\tau_0$ and $1/\tau_{x,y,z}=0$. According $z_{00}=\tau/\tau_0-\tau/\tau_z$, we find $z_{00}\approx 1$ in this case. Since $\ell_{0}^{-2}\propto(1/z_{00}-1)$, we find that $\ell_{0}$ become very large such as $\ell_{0}=1000$nm. In this case, the Cooperon contribution is quite large as can be seen from the figure. Since the magnetic field breaks the weak localization, the magneto-conductivity rapidly increases with the increasing magnetic field. One can see magneto-conductivity quickly become saturated to some constant values at about $B=0.5$T.

The panel (b) corresponds to the other limiting case where $\tau_{z}$ is close to $\tau_{0}$. It is easy to see that $z_{00}\approx0$ in this case. Thus $\ell_{0}^{-2}$ becomes divergent or $\ell_{0}$ is very small such as $\ell_{0}=8$nm. Form Eq.(\ref{eq:sig-B0}), one can see that the conductivity with quantum interference is proportional to $\ln\frac{\ell_{0}^{-2}+\ell_{e}^{-2}}{\ell_{0}^{-2}+\ell_{\phi}^{-2}}$ which become very small for divergent $\ell_0^{-2}$. Form this, it is clear that the Cooperon correction is suppressed in this case. It can be see from the figure since the overall scale of panel (b) is much smaller than panel (a). In panel (b) of \cref{fig:MC}, we can also see that the magneto-conductivity almost stays at zero until $B\approx 0.5$ T.

All the above results are obtained for $T=0$. Now we briefly discuss some possible temperature dependence. In the quantum diffusive regime with small $\ell_0^{-2}$, the weak localization correction of conductivities is proportional to $\ln\left(\ell_{\phi}/\ell_{e}\right)$, where the phase coherence length $\ell_{\phi}$ has a temperature dependence of $T^{-p/2}$ with parameter $p$ depending on the decoherence mechanisms. Electron-electron ($p=1$) and electron-phonon ($p=3$) interaction are two dominant decoherence mechanisms in two-dimensional disordered metals \cite{Akkermans2007:Cam,Datta1995:Cam}. At sufficiently low temperature, the phonon excitations are suppressed and the electron-electron interaction dominates the decoherence effect. Therefore, the temperature dependence of conductivities is mainly generated by $\ell_{\phi} \propto T^{-1/2}$, which can be verified in magneto-conductivity measurements.

\section{Anomalous Hall Conductivity}
\label{sec:AHC}

After the investigation of the longitudinal conductivity, we now turn to the study of the transverse component, which is usually considered under the name of anomalous Hall effect. The anomalous Hall effect originates from the interplay between the spin-orbit coupling and time-reversal symmetry breaking \cite{Nagaosa2010:RMP},
and can be classified into intrinsic and extrinsic contributions. The former comes from the Berry curvature of occupied bands below the Fermi surface, and the latter is the consequence of impurity scattering near the Fermi surface.
The weak localization behavior reveals that the Berry phase of the bulk states are zero, 
and thus the intrinsic anomalous Hall conductivity is always zero. This has also be verified 
in our numerical simulation, where a lattice version of Hamiltonian Eq.(II.1) can be obtained
by naive replacements $k_{x}\to\sin k_{x}$ and $k_{y}^{2}\to2\left(1-\cos k_{y}\right)$.
However, the doped impurities may result in non-zero anomalous Hall conductivity.

\begin{figure}[H]
\centering
\includegraphics[width=10cm]{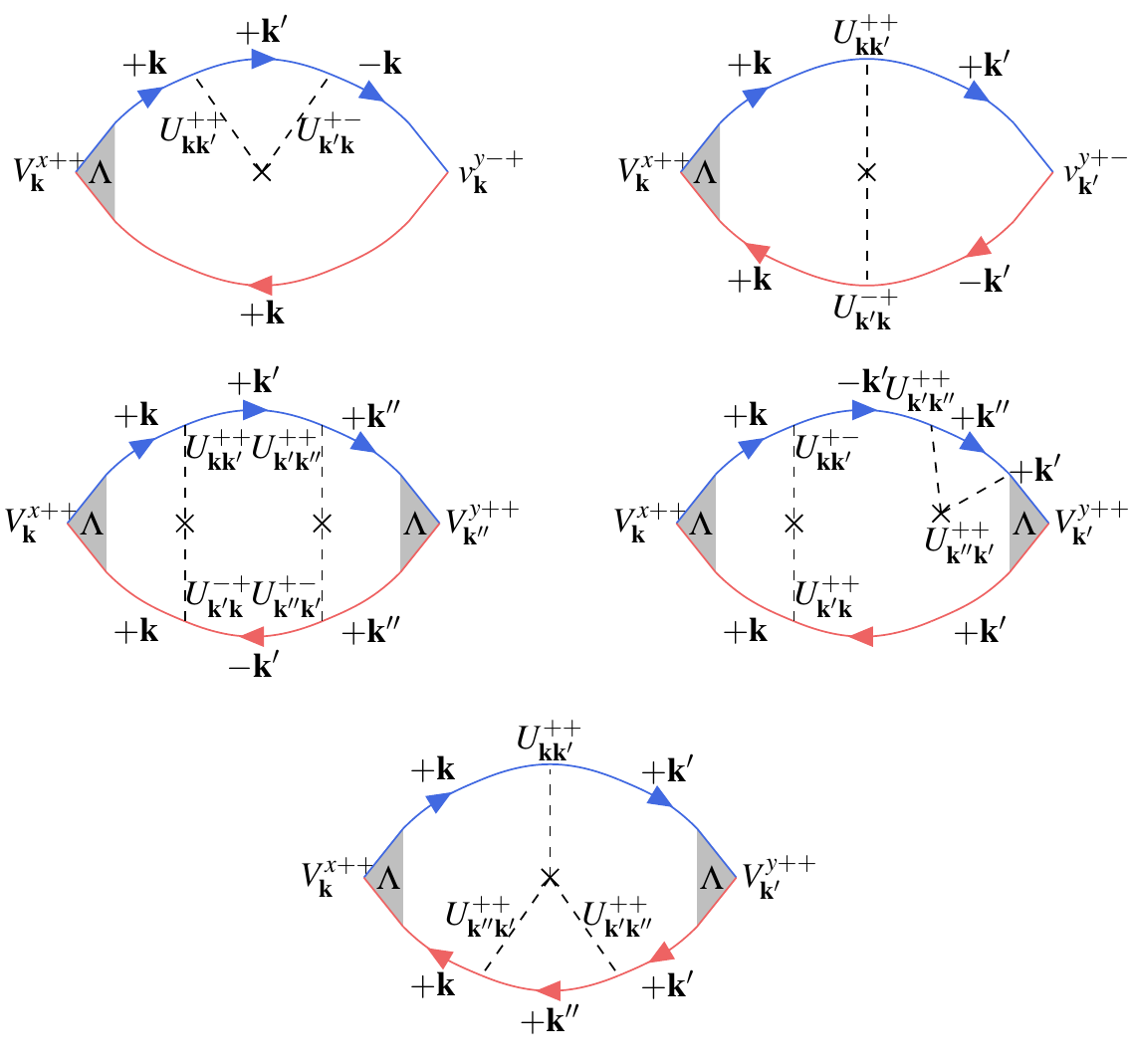}
\caption{Feynman diagrams for the extrinsic Hall conductivity calculation.
Here we list five representative diagrams.
Top four: Side-jump contribution. Bottom: Skew-scattering contribution.}
\label{fig:AHC}
\end{figure}

The calculation of extrinsic anomalous Hall conductivity is attributed to five representative Feynman diagrams\cite{Sinitsyn2007:PRB,Nagaosa2010:RMP,Yang2011:PRB,Lu2013:PRB} in \cref{fig:AHC}. One can exchange the position
of impurity scattering correlation lines and the velocity operators to obtain the full 16 possible diagrams. However, those
diagrams obtained by exchanging the lines are usually complex conjugate of the original diagrams. Therefore, we only need to calculate the contributions of these 5 representative diagrams and take the real part of them.

For later convenience, we slightly extend our previous notations to include both occupied and un-occupied bands.
The energy eigenvalues are $\epsilon_{s}(\mathbf{k})=s\sqrt{\lambda^{2}k_{x}^{2}+k_{y}^{4}}$,
with $s=+$ for conduction band and $s=-$ for valence band, respectively. And the corresponding
Bloch wavefunctions are
\begin{equation}\begin{aligned}
\left\vert \mathbf{k}s\right\rangle =
\frac{1}{\sqrt{2}}\left[\begin{array}{c}
s\\
\xi\mathrm{e}^{\mathrm{i}\phi}
\end{array}\right] .
\end{aligned}\end{equation}
The Born scattering amplitude is
\begin{equation}\begin{aligned}
U_{\mathbf{k}\mathbf{k}'}^{ss'}=\int\dfrac{\mathrm{d}^{2}\mathbf{r}}{S}\left\langle \mathbf{k}s\right\vert U(\mathbf{r})\left\vert \mathbf{k}'s'\right\rangle \mathrm{e}^{-\mathrm{i}(\mathbf{k}-\mathbf{k}')\mathbf{r}} ,
\end{aligned}\end{equation}
Then our previous defined amplitude is just one component as $U_{\mathbf{k}\mathbf{k}'}=U_{\mathbf{k}\mathbf{k}'}^{++}$.
The bare velocity operator is also extended to be $v_{\mathbf{k}}^{\mu ss'}=\left\langle \mathbf{k}s\right\vert \frac{\partial H}{\partial k_\mu}\left\vert \mathbf{k}s'\right\rangle$,
and likewise, the previous velocity is also one component $v_{\mathbf{k}}^{\mu}=v_{\mathbf{k}}^{\mu++}$.
Finally, the bare Green's function in \cref{eq:MGF} is also extended to include two components as
\begin{equation}\begin{aligned}
G_{s}(\mathbf{k},\mathrm{i}\omega_{m})=\frac{1}{\mathrm{i}\omega_{m}-\epsilon_{s\mathbf{k}}+\mathrm{i}/(2\tau)\sgn\omega_{m}}
\label{eq:MGFs}
\end{aligned}\end{equation}

\subsection{Side-jump Mechanism}

The side-jump contribution consists of diagrams in \cref{fig:AHC} (top left $\sigma_{xy}^{\mathrm{sj}1}$,
top right $\sigma_{xy}^{\mathrm{sj}2}$, middle left $\sigma_{xy}^{\mathrm{sj}3}$, middle right $\sigma_{xy}^{\mathrm{sj}4}$),
and can be calculated as
\begin{equation}\begin{aligned}
\sigma_{xy}^{\mathrm{sj}1}=&\frac{1}{\beta\nu_{n}}\sum_{m}\int\frac{\mathrm{d}^{2}\mathbf{k}}{\left(2\pi\right)^{2}}\int\frac{\mathrm{d}^{2}\mathbf{k}'}{\left(2\pi\right)^{2}}V_{\mathbf{k}}^{x++}G_{+}(\mathbf{k},\mathrm{i}\omega_{m})\left\langle U_{\mathbf{k}\mathbf{k}'}^{++}U_{\mathbf{k}'\mathbf{k}}^{+-}\right\rangle_{\mathrm{imp}} \\
&\times G_{+}(\mathbf{k}',\mathrm{i}\omega_{m})G_{-}(\mathbf{k},\mathrm{i}\omega_{m})v_{\mathbf{k}}^{y-+}G_{+}(\mathbf{k},\mathrm{i}\omega_{m}+\mathrm{i}\nu_{n}) \\
\sigma_{xy}^{\mathrm{sj}2}=&\frac{1}{\beta\nu_{n}}\sum_{m}\int\frac{\mathrm{d}^{2}\mathbf{k}}{\left(2\pi\right)^{2}}\int\frac{\mathrm{d}^{2}\mathbf{k}'}{\left(2\pi\right)^{2}}V_{\mathbf{k}}^{x++}G_{+}(\mathbf{k},\mathrm{i}\omega_{m})\left\langle U_{\mathbf{k}\mathbf{k}'}^{++}U_{\mathbf{k}'\mathbf{k}}^{-+}\right\rangle_{\mathrm{imp}} \\
&\times G_{+}(\mathbf{k}',\mathrm{i}\omega_{m})v_{\mathbf{k}'}^{y+-}G_{-}(\mathbf{k}',\mathrm{i}\omega_{m}+\mathrm{i}\nu_{n})G_{+}(\mathbf{k},\mathrm{i}\omega_{m}+\mathrm{i}\nu_{n}) \\
\sigma_{xy}^{\mathrm{sj}3}=&\frac{1}{\beta\nu_{n}}\sum_{m}\int\frac{\mathrm{d}^{2}\mathbf{k}}{\left(2\pi\right)^{2}}\int\frac{\mathrm{d}^{2}\mathbf{k}'}{\left(2\pi\right)^{2}}\int\frac{\mathrm{d}^{2}\mathbf{k}''}{\left(2\pi\right)^{2}}V_{\mathbf{k}}^{x++}G_{+}(\mathbf{k},\mathrm{i}\omega_{m})\left\langle U_{\mathbf{k}\mathbf{k}'}^{++}U_{\mathbf{k}'\mathbf{k}}^{-+}\right\rangle_{\mathrm{imp}}G_{+}(\mathbf{k}',\mathrm{i}\omega_{m})\left\langle U_{\mathbf{k}'\mathbf{k}''}^{++}U_{\mathbf{k}''\mathbf{k}'}^{+-}\right\rangle_{\mathrm{imp}} \\
&\times G_{+}(\mathbf{k}'',\mathrm{i}\omega_{m})V_{\mathbf{k}''}^{y++}G_{+}(\mathbf{k}'',\mathrm{i}\omega_{m}+\mathrm{i}\nu_{n})G_{-}(\mathbf{k}',\mathrm{i}\omega_{m}+\mathrm{i}\nu_{n})G_{+}(\mathbf{k},\mathrm{i}\omega_{m}+\mathrm{i}\nu_{n}) \\
\sigma_{xy}^{\mathrm{sj}4}=&\frac{1}{\beta\nu_{n}}\sum_{m}\int\frac{\mathrm{d}^{2}\mathbf{k}}{\left(2\pi\right)^{2}}\int\frac{\mathrm{d}^{2}\mathbf{k}'}{\left(2\pi\right)^{2}}\int\frac{\mathrm{d}^{2}\mathbf{k}''}{\left(2\pi\right)^{2}}V_{\mathbf{k}}^{x++}G_{+}(\mathbf{k},\mathrm{i}\omega_{m})\left\langle U_{\mathbf{k}\mathbf{k}'}^{+-}U_{\mathbf{k}'\mathbf{k}}^{++}\right\rangle_{\mathrm{imp}}G_{-}(\mathbf{k}',\mathrm{i}\omega_{m})\left\langle U_{\mathbf{k}'\mathbf{k}''}^{-+}U_{\mathbf{k}''\mathbf{k}'}^{++}\right\rangle_{\mathrm{imp}} \\
&\times G_{+}(\mathbf{k}'',\mathrm{i}\omega_{m})G_{+}(\mathbf{k}',\mathrm{i}\omega_{m})V_{\mathbf{k}'}^{y++}G_{+}(\mathbf{k}',\mathrm{i}\omega_{m}+\mathrm{i}\nu_{n})G_{+}(\mathbf{k},\mathrm{i}\omega_{m}+\mathrm{i}\nu_{n})
\end{aligned}\end{equation}
And after some integration we come to the fact that the contribution of side-jump mechanism is zero. The reason for vanishing side-jump contribution is because the energy band is gapless.

\subsection{Skew-scattering Mechanism}

In the skew-scattering mechanism, the Hall conductivity comes from two diagrams of third order correction.
One diagram is shown in the bottom panel of \cref{fig:AHC} with inverse "Y"-type disorder correlation lines, and the other diagram is complex conjugate of previous one. Combining these two diagrams, the Hall conductivity from skew-scattering is
\begin{equation}\begin{aligned}
\sigma_{xy}^{\mathrm{sk}}=2\re& \frac{1}{\beta\nu_{n}}\sum_{m}\int\frac{\mathrm{d}^{2}\mathbf{k}}{\left(2\pi\right)^{2}}\int\frac{\mathrm{d}^{2}\mathbf{k}'}{\left(2\pi\right)^{2}}\int\frac{\mathrm{d}^{2}\mathbf{k}''}{\left(2\pi\right)^{2}}V_{\mathbf{k}}^{x++}G_{+}(\mathbf{k},\mathrm{i}\omega_{m})\left\langle U_{\mathbf{k}\mathbf{k}'}^{++}U_{\mathbf{k}'\mathbf{k}''}^{++}U_{\mathbf{k}''\mathbf{k}}^{++}\right\rangle_{\mathrm{imp}} \\
&\times G_{+}(\mathbf{k}',\mathrm{i}\omega_{m})V_{\mathbf{k'}}^{y++}G_{+}(\mathbf{k}',\mathrm{i}\omega_{m}+\mathrm{i}\nu_{n})G_{+}(\mathbf{k}'',\mathrm{i}\omega_{m}+\mathrm{i}\nu_{n})G_{+}(\mathbf{k},\mathrm{i}\omega_{m}+\mathrm{i}\nu_{n})
\end{aligned}\end{equation}
After the momentum summation, we find a non-zero result for the Hall conductivity as
\begin{equation}\begin{aligned}
\sigma_{xy}^{\mathrm{sk}}= -\frac{\pi^{3}}{6}\eta_{x}\eta_{y}\sqrt{\frac{5}{2\pi}D_{x}D_{y}}n_{z}\overline{u_{z}^{3}}\rho(\epsilon_{F})^{3}\tau
\label{eq:sig-sk}
\end{aligned}\end{equation}
Since the intrinsic and side-jump part are zero, the above result is the only contribution to the transverse conductivity.

\section{Universal Conductance Fluctuation}
\label{sec:UCF}

\begin{figure}
\centering
\includegraphics[width=8cm]{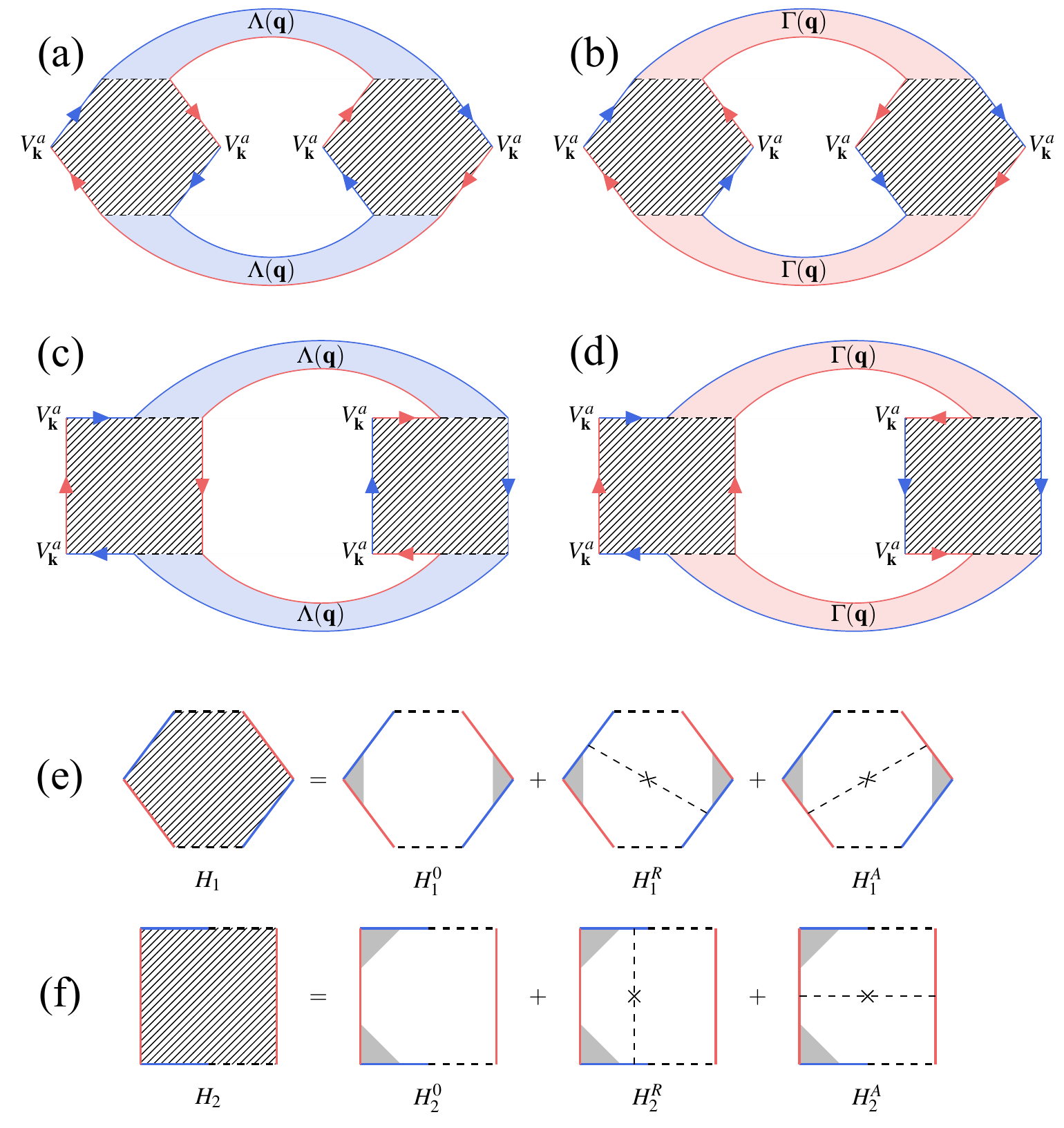}
\caption{Feynman diagrams (a-d) for the conductivity correlation function.
The blue and red stripes denote the diffusons and Cooperons, respectively.
The blocks rendered with lines are Hikami boxes in (e) and (f).
(a-b) are for fluctuations of diffusion constants and
(c-d) are for fluctuations of density of states.
(e-f) are bared and dressed Hikami boxes, and the gray parts are
vertex correction mentioned before.}
\label{fig:UCF}
\end{figure}

Universal conductance fluctuation (UCF) is another famous quantum interference phenomenon due to disorder that goes along
with weak localization/anti-localization in mesoscopic physics \cite{PALee1985:PRL,PALee1987:PRB,LiZG2015:Acta,Akkermans2007:Cam}.
The fluctuation of conductance, or here in 2D, the conductivity shows certain statistical distribution from sample to sample,
yet its root mean square is unrelated to impurity configuration and only slightly rely on the shape and dimension of the sample.
The weak localization/anti-localization correction is of magnitude $e^{2}/\hbar$ and so is the UCF.
Such an effect is sensitive to temperature. The temperature will affects not only the Fermi level but also the phase coherence length
$\ell_{\phi}$, and excites the thermal diffusive motion of electrons, characterized by $\ell_{T}=\sqrt{\hbar\beta D_{\mu}}$. Hence we
present the investigation of the UCF of semi-Dirac semimetal at zero temperature. The conductivity correlation function is
\begin{equation}\begin{aligned}
F_{\mu}(\epsilon-\epsilon')=\left\langle \delta\sigma_{\mu\mu}(\epsilon) \delta\sigma_{\mu\mu}(\epsilon')\right\rangle_{\mathrm{imp}} .
\end{aligned}\end{equation}
Here $\delta\sigma=\sigma-\left\langle \sigma\right\rangle $.
If one generalize the classical Einstein's relation to a random variable relation as $\sigma=\mathcal{D}\rho$, then the conductivity correlation can be separated into two parts
\begin{equation}\begin{aligned}
\left\langle \delta\sigma(\epsilon)\delta\sigma(\epsilon')\right\rangle = \left\langle \sigma\right\rangle^{2}
\left(\frac{\left\langle \delta\rho(\epsilon)\delta\rho(\epsilon')\right\rangle }{\left\langle \rho\right\rangle^{2}}+
\frac{\left\langle \delta\mathcal{D}(\epsilon)\delta\mathcal{D}(\epsilon')\right\rangle }{\left\langle \mathcal{D}\right\rangle^{2}}\right).
\label{eq:del-sig}
\end{aligned}\end{equation}
We can interpret the fluctuation as the joint contribution from the fluctuations of DOS and diffusion constants,
assuming the two fluctuations are uncorrelated. The explicit correspondence is shown in \cref{fig:UCF}.

The diagrams in \cref{fig:UCF} (a-d) can be calculated as the product of the building blocks such as the Hikami boxes and the diffuson/Cooperon vertex function. Let us first collect the expressions for all these building blocks as follows.
The bare diffuson is
\begin{equation}\begin{aligned}
\Lambda_{\mathbf{k}\mathbf{k}'}^{0}=\left\langle U_{\mathbf{k}\mathbf{k}'}U_{\mathbf{k}'\mathbf{k}}\right\rangle_{\mathrm{imp}}=
\frac{1}{2\pi\rho\tau S}\left[1+
\left(\frac{\tau}{\tau_{0}}-\frac{\tau}{\tau_{z}}\right)\cos\left(\phi-\phi'\right)\right]
\end{aligned}\end{equation}
and the Bethe-Salpeter equation for the diffuson is
\begin{equation}\begin{aligned}
\Lambda_{\mathbf{k}_{\alpha}\mathbf{k}_{\beta}}=\Lambda_{\mathbf{k}_{\alpha}\mathbf{k}_{\beta}}^{0}+
\sum_{\mathbf{k}_{\mu}}\Lambda_{\mathbf{k}_{\alpha}\mathbf{k}_{\mu}}^{0}
G\left(\mathbf{k}_{\mu},\mathrm{i}\omega_{m}\right)G\left(\mathbf{k}_{\mu}+\mathbf{q},\mathrm{i}\omega_{m}+\mathrm{i}\nu_{n}\right)
\Lambda_{\mathbf{k}_{\mu}\mathbf{k}_{\beta}}
\end{aligned}\end{equation}
Through a similar method we have used in solving the full Cooperon vertex function, we obtain the full diffuson vertex function as
\begin{equation}\begin{aligned}
\Lambda(\mathbf{q},\mathrm{i}\nu_{n})=\frac{1}{2\pi\rho\tau^{2}S}
\frac{1}{\nu_{n}+\left(\dfrac{\tau}{\tau_{0}}-\dfrac{\tau}{\tau_{z}}\right)\left(D_{x}q_{x}^{2}+D_{y}q_{y}^{2}\right)}
\end{aligned}\end{equation}
Similarly, the full Cooperon vertex function is
\begin{equation}\begin{aligned}
\Gamma(\mathbf{q},\mathrm{i}\nu_{n})=\frac{1}{2\pi\rho\tau^{2}S}\dfrac{1}{\nu_{n}+\left(\dfrac{1}{z_{00}}-1\right)\tau^{-1}+g_{x}D_{x}q_{x}^{2}+g_{y}D_{y}q_{y}^{2}}
\end{aligned}\end{equation}
For later convenience, we drop the prefactor of the above diffuson/Cooperon vertex and define the diffuson/Cooperon kernel as
\begin{equation}\begin{aligned}
P_{D}(\mathbf{q},\nu)=&\frac{1}{-\mathrm{i}\nu+\eta\left(D_{x}q_{x}^{2}+D_{y}q_{y}^{2}\right)} \\
P_{C}(\mathbf{q},\nu)=&\frac{1}{-\mathrm{i}\nu+\Omega_{0}+g_{x}D_{x}q_{x}^{2}+g_{y}D_{y}q_{y}^{2}}
\end{aligned}\end{equation}
Here we introduced the Cooperon gap $\Omega_{0}=\left(z_{00}^{-1}-1\right)\tau^{-1}$ and also $\eta=z_{00}$. We also have made the analytic continuation as $i\nu_n\to \nu$ in above expressions.

Then we consider the two types of Hikami boxes in \cref{fig:UCF} (e-f). For $H_{1}$, only the bare Hikami box $H_{1}^{0}$ is non-zero.
The other two Hikami boxes $H_{1}^{R/A}$ cancel out because the additional disorder correlation lines make the two incoming momenta uncorrelated.
Thus they vanish after the angular average of the incoming momenta. This is also true for $H_{2}^{A}$. Therefore, we find the summation of the first type of Hikami as
\begin{equation}\begin{aligned}
H_{1}=H_{1}^{0}=\frac{1}{\beta\nu_{n}}\sum_{m} \int\frac{\mathrm{d}^{2}\mathbf{k}}{(2\pi)^{2}} \left(V_{\mathbf{k}}^{\mu}\right)^{2}
G\left(\mathbf{k},\mathrm{i}\omega_{m}\right)^{2}G\left(\mathbf{k},\mathrm{i}\omega_{m}+\mathrm{i}\nu_{n}\right)^{2}
=\frac{4\pi\rho\eta_{\mu}^{2}D_{\mu}\tau^{-1}}{\left(\nu_{n}+\tau^{-1}\right)^{3}}
\end{aligned}\end{equation}
In the zero-temperature limit we arrived at $H_{1}=4\pi\rho \eta_{\mu}^{2}D_{\mu}\tau^{2}$. Similarly, for $H_{2}$ we have $H_{2}^{0}=H_{1}^{0}$, $H_2^A=0$
and the disorder dressed one is $H_{2}^{R}=\xi_\mu H_{2}^{0}$. Combining all the three terms, we find $H_{2}=4\pi\rho \eta_{\mu}^{2}\left(1+\xi_{\mu}\right)D_{\mu}\tau^{2}$.

Collect all the above results, the fluctuations of diffusion constants corresponding to the sum of diagrams of \cref{fig:UCF} (a-b) is
\begin{equation}\begin{aligned}
\overline{\delta\sigma_{\mu\mu}(\epsilon)\delta\sigma_{\mu\mu}(\epsilon')}^{(1)}=& \left(H_{1}\right)^{2} \sum_{\mathbf{q}}
\left[\left|\Lambda(\mathbf{q},\epsilon-\epsilon')\right|^{2}+\left|\Gamma(\mathbf{q},\epsilon-\epsilon')\right|^{2}\right] \\
=& 4\eta_{\mu}^{2}D_{\mu}^{2} /S^{2} \sum_{\mathbf{q}}
\left[\left|P_{D}(\mathbf{q},\epsilon-\epsilon')\right|^{2}+\left|P_{C}(\mathbf{q},\epsilon-\epsilon')\right|^{2}\right]
\end{aligned}\end{equation}

Similarly, the fluctuations of density of states corresponding to the sum of diagrams of \cref{fig:UCF} (c-d) together with their complex conjugate gives
\begin{equation}\begin{aligned}
\overline{\delta\sigma_{\mu\mu}(\epsilon)\delta\sigma_{\mu\mu}(\epsilon')}^{(2)}=& 2\left(H_{2}\right)^{2}
\sum_{\mathbf{q}}\re\left[\Lambda(\mathbf{q},\epsilon-\epsilon')^{2}+\Gamma(\mathbf{q},\epsilon-\epsilon')^{2}\right] \\
=& 8\eta_{\mu}^{2}\left(1+\xi_{\mu}\right)^{2}D_{\mu}^{2}/S^{2} \sum_{\mathbf{q}}
\re\left[P_{D}(\mathbf{q},\epsilon-\epsilon')^{2}+P_{C}(\mathbf{q},\epsilon-\epsilon')^{2}\right]
\end{aligned}\end{equation}

Plug the above two results into Eq.(\ref{eq:del-sig}), we will arrive at the final results of the conductance fluctuation.
Here we will apply this general formalism to two different regimes as follows

\subsection{Universal Conductance Fluctuation in mesoscopic regime}

 In the \textit{mesoscopic} regime, $\ell_{\phi}>L_{x,y}$ where $L_{x,y}$ is the sample size along $x,y$ direction.
The momentum is quantized as $\mathbf{q}_{\mu}=\pi n_{\mu}/L_{\mu}$. Applying this momentum quantization to Eq.(\ref{eq:del-sig}), we find the UCF as
\begin{equation}\begin{aligned}
\left\langle \delta\sigma_{\mu\mu}^{2}\right\rangle =& \frac{4\eta_{\mu}^{2}D_{\mu}^{2}\left(1+2\left(1+\xi_{\mu}\right)^{2}\right)}{\pi^{4}D_{x}D_{y}}
\sum'_{n_{x},n_{y}}\Bigg[\frac{1}{\eta^{2}\left(n_{x}^{2}\mathcal{R}_{1}\mathcal{R}_{2}^{-1/2}+n_{y}^{2}\mathcal{R}_{1}^{-1}\mathcal{R}_{2}^{1/2}\right)^{2}} \\
&+\frac{1}{\left(\frac{\Omega_{0}S}{\pi^{2}\sqrt{D_{x}D_{y}}}+g_{x}n_{x}^{2}\mathcal{R}_{1}\mathcal{R}_{2}^{-1/2}+g_{y}n_{y}^{2}\mathcal{R}_{1}^{-1}\mathcal{R}_{2}^{1/2}\right)^{2}}\Bigg],
\label{eq:UCFv1}
\end{aligned}\end{equation}
Here we introduced the geometrical ratio $\mathcal{R}_{1}=L_{y}/L_{x}$ and the anisotropic diffusion ratio $\mathcal{R}_{2}=D_{y}/D_{x}\sim\lambda^2/\epsilon_{F}$.
In the above formula, we only consider the case of $\epsilon=\epsilon'=\epsilon_{F}$ for simplicity.  The general case with non-zero Fermi energy difference can be achieved by applying a gate voltage in experiment. The prime above $\sum_{n_{x},n_{y}}$ means that the $n_x=0$ term is excluded, otherwise the denominator will vanish if one also have $n_y=0$ at the same time. The reason for this exclusion is because the conductance is always measured by attaching leads to the sample. Thus, we can, for example, assume that the leads are attached in the $x$ direction, corresponding to absorbing walls or the Dirichlet boundary condition where the $n_{x}=0$ mode is removed. On the other hand, in the $y$ direction, we have hard walls or the Neumann boundary conditions where the $n_{y}=0$ mode is included. Note that the first term inside the bracket of \cref{eq:UCFv1} comes from diffuson contribution and the other term is from Cooperon.

\begin{figure}
\centering
\subfigure{\begin{minipage}[t]{0.45\textwidth}
\centering
\includegraphics[scale=0.8]{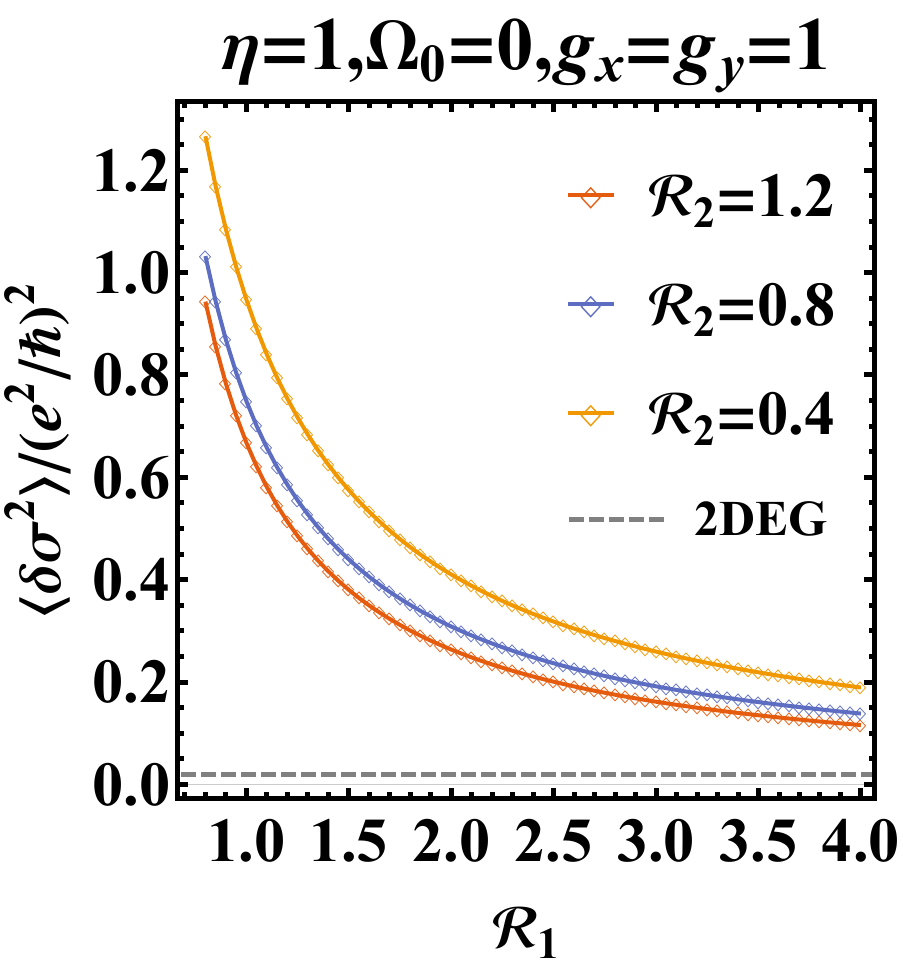}
\end{minipage}
}
\caption{The universal conductance fluctuation in the $x$ direction as a function of $\mathcal{R}_{1}$, for $\mathcal{R}_{2}=0.4,\,0.6,\,0.8$.
Other than the parameter shown in the figure, we also used $\eta_{x}=3$ and $\xi_{x}=-1/3$ which corresponds to the case with only $U_{0}(\mathbf{r})$ impurity potential. The grey dashed line shows the UCF of 2D electron gas.}
\label{fig:UCF1}
\end{figure}

By taking proper phenomenological parameters, we recover the result for two-dimensional electron gas (2DEG):
\begin{equation}\begin{aligned}
\left\langle \delta\sigma_{\mu\mu}^{2}\right\rangle = \frac{12}{\pi^{4}}\sum_{n_{x},n_{y}}\frac{1}{\left(n_{x}^{2}+n_{y}^{2}\right)^{2}}
\end{aligned}\end{equation}
which does not depend on $\mathcal{R}_{1,2}$. However, the UCF of \cref{eq:UCFv1} shows a dependence on the sample shape, impurity scattering type,
and intrinsic properties such as $\lambda^2/\epsilon_{F}$. In \cref{fig:UCF1}, we plot the conductance fluctuation of \cref{eq:UCFv1} as a function of $\mathcal{R}_{1}$ for several different $\mathcal{R}_{2}$. One can see the fluctuation approaches to the value of 2D electron gas as $\mathcal{R}_1$ increasing.
We would like to mention that the contribution from the Cooperon will be suppressed if a strong magnetic field is applied. In this case, only the contribution of the diffuson remains.

\subsection{Conductance Fluctuation in quantum diffusive regime}

In the opposite quantum diffusive regime, we have $\ell_{\phi}\lesssim L$. In this case, one can treat the momentum as continuous variables and carry out the momentum integration to find the following result of conductance fluctuation
\begin{equation}\begin{aligned}
\left\langle \delta\sigma_{\mu\mu}^{2}\right\rangle =\frac{\eta_{\mu}^{2}D_{\mu}^{2}\left[1+2\left(1+\xi_{\mu}\right)^{2}\right]}{\pi g_{x}g_{y}D_{x}D_{y}S}
\left(g_{x}g_{y}\frac{\ell_{\phi}^{2}-\ell_{e}^{2}}{\eta^{2}}+\frac{1}{\ell_{\phi}^{-2}+\ell_{0}^{-2}}-\frac{1}{\ell_{e}^{-2}+\ell_{0}^{-2}}\right)
\end{aligned}\end{equation}
Since now the conductance fluctuation depends on the $\ell_\phi$ and $\ell_e$, etc, it is not universal any more. This result can be compared with the conductivity of Eq.(\ref{eq:sig-qi}) which is also valid in the quantum diffusive regime.

Now we can consider the influence of magnetic field on the conductance fluctuation. Following the similar steps in the computation of magneto-conductivity, we can replace the momentum integration in the above formula by a summation of Landau level. Note that since the diffuson vertex $\Lambda_{\mathbf{k}\mathbf{k}'}$ depend on the difference of the two momentum, the vector potential of magnetic field will cancel out if one make a Peierls substitution. Because of this, only the Cooperon vertex will depend on the vector potential and requires to take account the Landau level summation. With all above considerations, we can find the conductance fluctuation under the magnetic field as
\begin{equation}\begin{aligned}
\left\langle \delta\sigma_{\mu\mu}^{2}(B)\right\rangle = & \frac{\eta_{\mu}^{2}D_{\mu}^{2}\left[1+2\left(1+\xi_{\mu}\right)^{2}\right]}{\pi g_{x}g_{y}D_{x}D_{y}S}
\Bigg\{  g_{x}g_{y}\frac{\ell_{\phi}^{2}-\ell_{e}^{2}}{\eta^{2}} \\  &  +\frac{4}{3}\chi^{\frac{4}{3}}\left[
\zeta_{\frac{7}{3}}\left(\chi\left(\ell_{\phi}^{-2}+\ell_{0}^{-2}\right)^{\frac{3}{4}}+
\frac{1}{2}\right)-\zeta_{\frac{7}{3}}\left(\chi\left(\ell_{\phi}^{-2}+\ell_{0}^{-2}\right)^{\frac{3}{4}}+\frac{1}{2}\right)\right] \Bigg\}
\end{aligned}\end{equation}
Here $\zeta_{\frac{7}{3}}(x)$ is the Hurwitz zeta function.

\begin{figure}
\centering
\subfigure{\begin{minipage}[t]{0.45\textwidth}
\centering
\includegraphics[scale=0.8]{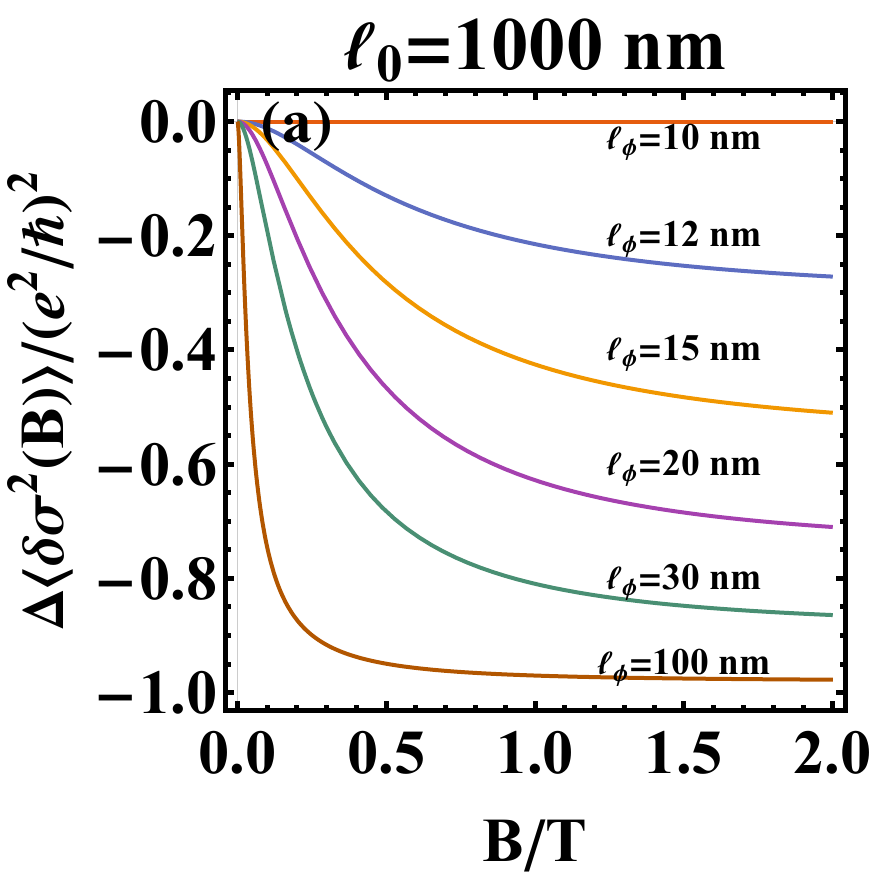}
\end{minipage}
}
\subfigure{\begin{minipage}[t]{0.45\textwidth}
\centering
\includegraphics[scale=0.8]{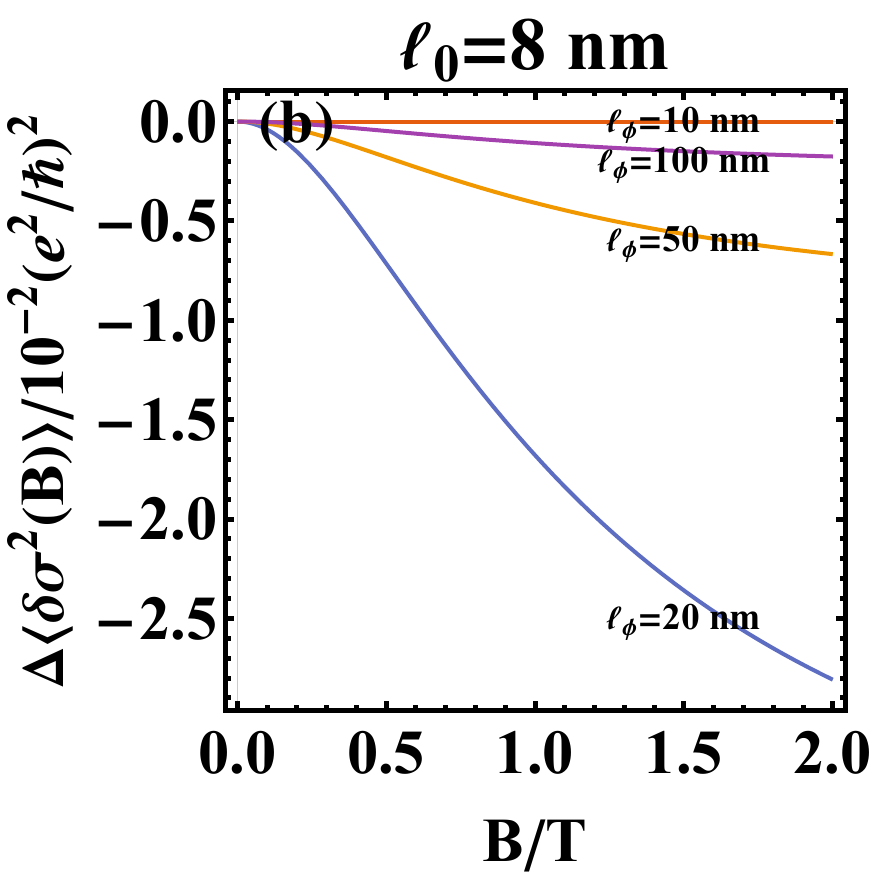}
\end{minipage}
}
\caption{The difference of conductance fluctuation $\Delta\left\langle\delta\sigma^{2}(B)\right\rangle$ as a function of the magnetic field. In panel (a), the adjustable $\ell_{0}=1000$nm and the phase coherence length $\ell_\phi=$10, 12, 20, 50, 200 nm from top to bottom. In panel (b), the adjustable $\ell_{0}=8$nm and the phase coherence length $\ell_\phi=$20, 50, 100, 10 nm from top to bottom. Other parameters are $\ell_{e}=10$ nm, $\eta=g_x=g_y=1$.}
\label{fig:UCF2}
\end{figure}

In order to visualize the above results, we introduce the difference of the conductance fluctuations with and without magnetic field as
\begin{equation}
\Delta\left\langle \delta\sigma_{\mu\mu}^{2}(B)\right\rangle
=\left\langle \delta\sigma_{\mu\mu}^{2}(B)\right\rangle-\left\langle \delta\sigma_{\mu\mu}^{2}(0)\right\rangle
\end{equation}
To be specific, we apply the above general formalism to a square shaped sample with the critical size $L=\ell_{\phi}$. In \cref{fig:UCF2}, the difference $\Delta\left\langle\delta\sigma^{2}(B)\right\rangle$ without the prefactor is plotted as a function of $B$ for two limiting cases with $\ell_0=1000$ nm and $\ell_0=8$ nm, which are the same as in \cref{fig:MC}. A series of values for $\ell_\phi$ are assumed in both panel (a) and (b).

To understand the behavior of this difference, we first note that the diffuson does not depend on the magnetic field, thus its contribution cancels out. We expect that $\Delta\left\langle\delta\sigma^{2}(B)\right\rangle$ only shows the difference of Cooperon contribution with and without magnetic field. In panel (a) with large $\ell_0$, the Cooperon effect is large in the absence of magnetic field, which generates a large positive conductance fluctuation. This fluctuation is suppressed by the increasing magnetic field. Therefore, in panel (a), we see that $\Delta\left\langle\delta\sigma^{2}(B)\right\rangle$ is large and negative. On the other hand, in panel (b) with small $\ell_0$, the Cooperon effect is small even without magnetic field. Because of this, one can see that the overall scale of panel (b) is two order of magnitudes smaller than panel (a).

\section{Conclusion and Remarks}
\label{sec:Conc}

In this work we studied the transport properties of the so-called semi-Dirac semimetals doped with various types of impurities. We use the diagrammatic perturbation method to obtain the zero-temperature electrical conductivities in both $x$ and $y$ direction. More specifically, we systematically explored the weak localization phenomenon and calculated the magneto-conductivity under a finite external magnetic field. Furthermore, the anomalous Hall conductivity generated by the side-jump mechanism and skew-scattering mechanism is investigated. It is found that the side-jump contribution is zero due to the vanishing energy gap. On the other hand, the third order perturbation of impurity potential $U_{z}(\mathbf{r})$
give rise to a non-zero skew-scattering contribution. At last, in the mesoscopic regime, we found that the UCF of semi-Dirac semimetal depend on the sample shape, impurity scattering type and some other intrinsic properties. In the quantum diffusive regime, the conductance fluctuation is suppressed by applying magnetic field.

We briefly discuss some possible extensions to our theory by taking other effects into consideration. We completely ignores the electron interaction in our discussion. It is known that the electron-electron interaction self energy $\Sigma_{\mathrm{ee}}(\mathbf{k},\mathrm{i}\omega_{m})$
can modify the DOS at the Fermi surface and induce additional conductivity correction. This is known as the Altshuler-Aronov effect \cite{Altshuler1985:Els}.
In the first order approximation of Dyson equation for full Green's function, the DOS shift is given by:
\begin{equation}\begin{aligned}
\delta\rho(\epsilon_{F})=-\frac{1}{\pi}\im\sum_{\mathbf{k}}
\left[G(\mathbf{k},\mathrm{i}\omega_{m})^{2}\Sigma_{\mathrm{ee}}(\mathbf{k},\mathrm{i}\omega_{m})\right]
\Big\vert_{\mathrm{i}\omega_{m}\to\epsilon_{F}+\mathrm{i}0^{+}},
\end{aligned}\end{equation}
and the resultant conductivity correction from the electron-electron interaction is found to be
$\sigma^{\mathrm{ee}}=\sigma^{\mathrm{sc}}\dfrac{\delta\rho(\epsilon_{F})}{\rho(\epsilon_{F})}$,
where $\sigma^{\mathrm{sc}}$ is the semi-classical Boltzmann conductivity aforementioned.

At last, we would like to emphasize that in the \cref{sec:MC}, the linear part of effective potential in \cref{eq:EigenEqs}
coming from the commutator is left out, which means that the original asymmetric effective potential is approximated
by a symmetric quartic potential. Because of this, the difference between the two components of eigen-wavefunction is eliminated. The validity of such an approximation is verified by numerical calculations and is proved to be very successful in \cite{Dietl2008:PRL}. However, the eigen-function of the lowest Landau level is fully polarized and only the lower component is non-zero. Therefore, in the ultra-quantum limit where only the lowest Landau level is partially filled (which can be achieved when the magnetic field is very strong), such an approximation may be invalid and needs further consideration. In this case, the magneto-conductivity may have a different dependence on the external magnetic field. All the possibilities discussed above are remained to be investigated in the future.

For the convenience of readers, we summarize the notations used throughout this paper 
and compare some quantities in 2DEG, 2D S-DSM, and 2D massless Dirac fermion in the Appendix \cref{tab:notations,tab:comp}.

\acknowledgments

This work is supported by NSFC under Grant No. 11874272 and Science Specialty Program of Sichuan University under Grant No. 2020SCUNL210.

\appendix
\section{Summary of notations}

\begin{table}[H]
  \centering
    \begin{tabular}{|c|c|c|}
    \hline
    \multicolumn{1}{|c|}{Notations} & \multicolumn{1}{c|}{Meaning} & \multicolumn{1}{c|}{Value} \\
    \hline
    $\tau_{0,x,y,z}$                & Relaxation time              & $\pi\rho(\epsilon_{F}) n_{0,x,y,z}\overline{u_{0,x,y,z}^{2}}$ \\
    \hline
    \multirow{2}[4]{*}{$D_{x,y}$}   &  \multirow{2}[4]{*}{Bare diffusion constants}  & $D_{x}=\dfrac{2}{3}\lambda^{2}\tau$ \\
    \cline{3-3}                     &                                                & $D_{y}=\dfrac{6\pi}{5K(1/2)^{2}}\left|\epsilon_{F}\right|\tau$ \\
    \hline
    \multirow{2}[4]{*}{$\eta_{x,y}$}& \multirow{2}[4]{*}{Vertex correction coefficients} & $\eta_{x}^{-1}=1-\dfrac{2}{3}\left(\dfrac{\tau}{\tau_{0}}-\dfrac{\tau}{\tau_{z}}\right)$ \\
    \cline{3-3}                     &                                                    & $\eta_{y}^{-1}= 1-\dfrac{5\pi}{48}\left(\dfrac{\tau}{\tau_{0}}-\dfrac{\tau}{\tau_{z}}\right)$ \\
    \hline
    \multirow{2}[4]{*}{$\xi_{x,y}$} & \multirow{2}[4]{*}{Ratio between dressed}          & $\xi_{x}=-\dfrac{1}{3}\left(\dfrac{\tau}{\tau_{0}}-\dfrac{\tau}{\tau_{z}}-2\dfrac{\tau}{\tau_{x}}\right)$  \\
    \cline{3-3}                     & and bare Hikami boxes                              & $\xi_{y}=-\dfrac{5\pi}{96}\left(\dfrac{\tau}{\tau_{0}}+\dfrac{\tau}{\tau_{z}}\right)$  \\
    \hline
    $\eta$                          & \makecell[c]{Correction coefficients in\\
                                                   diffuson vertex function}             & $\eta=\dfrac{\tau}{\tau_{0}}-\dfrac{\tau}{\tau_{z}}$ \\
    \hline
    $\Omega_{0}$                    &  Cooperon gap                                      & $\Omega_{0}=\left(\eta^{-1}-1\right)\tau^{-1}$ \\
    \hline
    $g_{x,y}$                       & \makecell[c]{Correction coefficients in\\
                                                   Cooperon vertex function}             & See  Eq.(\ref{eq:Z00})\\
    \hline
    $\ell_{0}$                      &  Length parameter                                  & $\ell_{0}^{-2}=\left(\eta^{-1}-1\right)\left(g_{x}D_{x}\tau\right)^{-1}$ \\
    \hline
    $\ell_{e/\phi}$                 & \makecell[c]{Mean free path/ \\
                                                   Phase coherence length}               & \makecell[c]{Extracted from experiments. \\ $\ell_{\phi}\sim T^{-p/2}$} \\
    \hline
    $\ell_{B}$                      & Magnetic length                                    & $\ell_{B}=\sqrt{\hbar/eB}$ \\
    \hline
    $\sigma_{\mu\mu}$               & Boltzmann conductivity                             & $\sigma_{\mu\mu}=\eta_{\mu}D_{\mu}\rho(\epsilon_{F})$ \\
    \hline
    $\sigma_{\mu\mu}^{0,R/A}$       & Bare and dressed Hikami boxes                      & See  Eq.(\ref{sig-H0}),(\ref{sig-H1})\\
    \hline
    $\sigma_{\mu\mu}^{\mathrm{qi}}$ & Quantum interference correction                    & See Eq.(\ref{eq:sig-B0})\\
    \hline
    $\sigma_{xy}^{\mathrm{sj}}$     & Side-jump contribution                             & $0$ \\
    \hline
    $\sigma_{xy}^{\mathrm{sk}}$     & Skew-scattering mechanism                          & See Eq.(\ref{eq:sig-sk})\\
    \hline
    $\Lambda(\mathbf{q},\mathrm{i}\nu_{n})$ & diffuson vertex function & $\Lambda(\mathbf{q},\mathrm{i}\nu_{n})=\dfrac{1}{2\pi\rho\tau^{2}S}\dfrac{1}{\nu_{n}+\eta\left(D_{x}q_{x}^{2}+D_{y}q_{y}^{2}\right)}$ \\
    \hline
    $\Gamma(\mathbf{q},\mathrm{i}\nu_{n})$  & Cooperon vertex function & $\Gamma(\mathbf{q},\mathrm{i}\nu_{n})=\dfrac{1}{2\pi\rho\tau^{2}S}\dfrac{1}{\nu_{n}+\Omega_{0}+g_{x}D_{x}q_{x}^{2}+g_{y}D_{y}q_{y}^{2}}$ \\
    \hline
    $P_{D}(\mathbf{q},\nu)$                 & diffuson kernel          & $P_{D}(\mathbf{q},\nu)=\dfrac{1}{-\mathrm{i}\nu+\eta\left(D_{x}q_{x}^{2}+D_{y}q_{y}^{2}\right)}$ \\
    \hline
    $P_{C}(\mathbf{q},\nu)$                 & Cooperon kernel          & $P_{C}(\mathbf{q},\nu)=\dfrac{1}{-\mathrm{i}\nu+\Omega_{0}+g_{x}D_{x}q_{x}^{2}+g_{y}D_{y}q_{y}^{2}}$ \\
    \hline
    \end{tabular}%
  \caption{Summary of notations.}
  \label{tab:notations}
\end{table}

\begin{table}[H]
  \centering
    \begin{tabular}{|c|c|c|c|}
    \hline
    \multicolumn{1}{|c|}{} & \multicolumn{1}{c|}{2DEG} & \multicolumn{1}{c|}{S-DSM} & \multicolumn{1}{c|}{2D massless Dirac fermion} \\
    \hline
    $\epsilon(\mathbf{k})$ & $\dfrac{k^{2}}{2m}$       & $\sqrt{\lambda^{2}k_{x}^{2}+k_{y}^{4}}$ & $\lambda k$ \\
    \hline
    $\rho(\epsilon)$ & $\dfrac{m}{2\pi}$ & $\dfrac{K(1/2)}{\sqrt{2}\lambda\pi^{2}}\epsilon^{1/2}$ & $\dfrac{\epsilon}{2\pi\lambda^{2}}$ \\
    \hline 
    \multirow{2}[4]{*}{$D_{x,y}$}& \multirow{2}[4]{*}{$\dfrac{1}{2}v_{F}^{2}\tau$} & $D_{x}=\dfrac{2}{3}\lambda^{2}\tau$ & \multirow{2}[4]{*}{$\dfrac{1}{2}\lambda^{2}\tau$} \\
    \cline{3-3}                  &                                                 & $D_{y}=\dfrac{6\pi}{5K(1/2)^{2}}\left|\epsilon_{F}\right|\tau$ &   \\
    \hline
    \multirow{2}[4]{*}{$\eta_{x,y}$}& \multirow{2}[4]{*}{1} & $\eta_{x} = 3$                                    & \multirow{2}[4]{*}{2} \\
    \cline{3-3}                     &                       & $\eta_{y} = \left(1-\dfrac{5\pi}{48}\right)^{-1}$ &   \\
    \hline
    \multirow{2}[4]{*}{$\xi_{x,y}$}& \multirow{2}[4]{*}{0} & $\xi_{x} = -1/3$     & \multirow{2}[4]{*}{$-1/4$} \\
    \cline{3-3}                    &                       & $\xi_{y} = -5\pi/96$ &   \\
    \hline
    $g_{x,y}$              & 1                         & 1                          & \\
    \hline
    $\eta$                 & 1                         & 1                          & 1 \\
    \hline
    $\Omega_{0}$           & 0                         & 0                          & $\tau^{-1}$ \\
    \hline
    \end{tabular}%
  \caption{Comparison between the 2DEG, S-DSM, and 2D massless Dirac fermion. The parameters are computed in the existence of $U_{0}(\mathbf{r})$ only.}
  \label{tab:comp}
\end{table}

\bibliographystyle{apsrev}
\bibliography{ref}
\end{document}